\documentclass[epj]{svjour}
\pdfoutput=1
\usepackage{graphics}
\usepackage{ulem}    
\usepackage{hyperref}

\begin{document}
\newcommand{\GeM}[1]{\mbox{Ge$_{#1}$}}
\newcommand{\AuGe}[2]{$\mathrm{Au_{#1}Ge_{#2}}$}

\newcommand{\AuNGeE}{$\mathrm{Au_nGe_{12}}$}

\newcommand{\beginsupplement}{%
        \setcounter{table}{0}
        \renewcommand{\thetable}{S\arabic{table}}%
        \setcounter{figure}{0}
        \renewcommand{\thefigure}{S\arabic{figure}}%
     }

\bibliographystyle{epj}
\authorrunning{McDermott and Newman}
\titlerunning{Stability of $\mathrm{Au_nGe_{m}}$  Clusters}
\title{Wade's Rules and the Stability of $\mathrm{Au_nGe_{m}}$  Clusters}
\author{Danielle McDermott\inst{1}\thanks{email: mcdermod@wabash.edu} 
and Kathie E. Newman\inst{2}\thanks{email: newman@nd.edu}
}                     
\institute{Department of Physics, Wabash College, Crawfordsville, IN 47933 \and Department of Physics, University of Notre Dame, Notre Dame IN 46556}
\date{Received: date / Revised version: date}
%

\abstract{ 
The properties of clusters 
formed from two connected $\mathrm{Ge_{m}}$~cage-like clusters, 
such as experimentally synthesized $\mathrm{Au_3Ge_{18}^{5-}}$, 
are examined using first-principles DFT methods.  
We focus particularly on $\mathrm{Au_nGe_{12}^{q-}}$ formed from a 
Wade-rules stable $\mathrm{Ge_{6}}$ cluster, where $n=0-3$ and $q=0,2$.  
The geometries, electronic structure, and thermal excitations 
of these clusters 
are examined using the SIESTA code.  
Cluster stability is tested using short molecular dynamics simulations.  
We find that intercluster bridges between \GeM{m} cages, 
formed of either Ge-Ge or Au-Ge bonds, 
can either bind a cluster together or tear it apart 
depending on the orientation of the bridging atoms with respect to the cages.  
The properties of 
neutrally charged 
$\mathrm{AuGe_{12}}$ and $\mathrm{Au_2Ge_{12}}$ are characterized, 
and we observe 
that radially directed molecular orbitals 
stabilize $\mathrm{AuGe_{12}}$ 
while a geometric asymmetry stabilizes 
$\mathrm{Au_2Ge_{12}}$ and $\mathrm{Au_3Ge_{18}}$.  
A two-dimensional $_2^\infty[\mathrm{Au_2Ge_{6}}]$ structure
is examined and found to be 
more stable than other periodic $[\mathrm{Au_nGe_{6}}]$ subunits.  
While no stable neutral isomers of $\mathrm{Au_3Ge_{12}}$ 
are observed in our calculations, 
our work suggests 
additional charge stabilizes isomers 
of both $\mathrm{Au_2Ge_{12}}$ and $\mathrm{Au_3Ge_{12}}$. 
} 
%

\maketitle
\section{Introduction}
\label{intro}

The element gold crystallizes 
in a face-centered cubic (fcc) structure, space group $Fm3m$, 
and germanium crystallizes in a diamond structure, space group $Fd\bar{3}m$.
Gold is a metal, with delocalized electrons, 
and germanium is a semiconductor, with its bonding electrons
localized in covalent bonds.
The story of the structure of these elements does not end here.  
In bulk form, 
synthetic chemists have grown several exotic crystalline forms of Ge, 
including a clathrate-II form with the same space group 
as the familiar semiconductor Ge, 
but having a 136-atom basis \cite{Guloy2006,Fassler2007}.
A microcrystalline allotrope of Ge is found, {\it allo-\/}Ge\cite{Kiefer2011},  
by removing Li from a compound Li$_7$Ge$_{12}$.
{\it allo-\/}Ge has a complex orthorhombic structure with 5- and 7-atom rings, 
space group $Pmc2_1$.  
Upon heating, {\it allo-\/}Ge changes form, 
becoming a hexagonal form of Ge, 4H-Ge, space
group $P6_3mc$, and then transforming to the standard diamond form, 
$\alpha$-Ge \cite{Gruttner1982}.  
The thermochemistry of these Ge allotropes have recently 
been investigated experimentally \cite{Zaikina2014}.
Allotropes of Ge are also found by varying pressure, 
e.g., $\beta-$Ge  (space group $I4_1/amd$), 
$\gamma-$Ge or ST12-Ge (space group $/P4_32_12$), 
and $\delta-$Ge or 
BC8-Ge (space group $Ia\bar{3}$) \cite{Fassler2007,Mujica2003}.
The bonding in these structures is particularly intriguing 
since in many cases it is not simply the standard $sp^3$ 
bonding one finds in group-IV and groups III-V semiconductors.

Gold's story is equally complex, 
with one interesting aspect being in the evolution of 
nanostructures of Au with $N$ atoms to the bulk form.
There is evidence that for small values of $N$, 
clusters with icosahedral 
and other interesting geometric shapes will form \cite{Koga2004,Baletto2005}.  
Nanoclusters of Au has been often been 
studied theoretically \cite{Wang2004,Kuo2005,Barnard2006} 
and ``phase'' diagrams of Au 
as a function of $N$ have been proposed \cite{Barnard2009}. 
Structural phase transition of Au under a variety of stresses have 
also been studied theoretically \cite{Durandurdu2007}.

It is know that the elements Au and Ge can combine in bulk form, 
given another element to provide 
spacing between groups of Au and Ge atoms.  
For example, there is a series of [AuGe] polyanion compounds 
where a six-fold ring of alternating Au and Ge forms, 
and then, dependent on ionic radius of the spacing rare earth ion (Cd, Lu, Sc), the ring is flat, becomes more puckered,
and bond angles evolve from ones characteristic of $sp^2$ 
bonding to those for $sp^3$ bonding \cite{Pottgen1996}.  
In another example, 
octahedra of Au$_5$Ge are found in a new compound SrAu$_3$Ge \cite{Lin2012}.
Au is often used in catalyzing reactions in Ge, e.g.,  
recently Au was used to self assemble
Ge micropatterns on Ge wafers \cite{Wu2008}.  
Fifty years ago, 
nanoparticles of Au were first used in the Vapor-Liquid-Solid technique
to grow nanowires of Si \cite{Wagner1964}.
The growth of Au-catalyzed Ge-nanowires is often studied; it appears that
Ge is ``fed'' from solution or vapor into the Au nanoparticle, with the nanowire being extruded from the base \cite{Wang2002}.
And while the binary Au-Ge system 
has a eutectic phase diagram \cite{Okamoto1984},
it has been shown that nanowires can be grown 
below the eutectic temperature \cite{Kodambaka2007}.
Dayeh {\it et al.} \cite{Dayeh2010} 
found Ge nanowire growth to be dependent on the diameter of the wires
and related this to the Gibbs-Thomson effect.
TEM images show surface melting of 
the crystalline Au-Ge alloy \cite{Sutter2008},
and more recently, the growth process has been studied 
using in-situ video rate TEM \cite{Gamalski2011}.
Most recently, Gamalski {\it et al.} \cite{Gamalski2012} observed metastable 
forms of AuGe during the nanowire growth process.  
The growth pathways of Ge emerging from Au, 
as studied recently \cite{Kim2012}, represent a fascinating
challenge for theorists, with some progress being made using
semi-empirical models \cite{Ryu2009,Dongare2009}
and simple kinetic Monte Carlo models of the growth 
of a ``cartoon'' nanowire \cite  {Reyes2013}.
And from another perspective, almost 25 years ago, 
it was observed that upon heating, 
a layer of Au grown on amorphous Ge would pass through the material, 
crystallizing it \cite {Tan1992}.
It appears that an atomic-scale description of Au-Ge bonds 
is essential for a broader physical understanding of these kinetic processes.  
We find the nanowire problem to be particularly fascinating, 
but unless one works with phenomenological 
or nearly phenomenological models of bonding, 
one cannot hope to study the numbers of atoms 
that must be involved in the process of nucleating nanowire growth.  

Our approach is to study the electronic structure of small clusters of Ge 
or Ge combined with Au.  
The tool that we utilize in our study is 
a first principles density functional theory (DFT) implemented 
with the SIESTA 
(Spanish Initiative for Electronic Simulations with Thousands of Atoms) 
code \cite{Sanchez-Portal1997,Soler2002}.  
This allows us to investigate
at an electronic level how a gold environment 
will effect the bonding of germanium in a cluster.  
A goal is to find stabilizing trends and 
thus identify possible building blocks for new compounds of Au with Ge.  
We investigate in particular cages of six or nine Ge atoms, 
including how they link together with and without the presence of Au atoms.

Ge clusters have previously been studied 
theoretically as a function of number 
(e.g., $n=5 - 10$ \cite{Li2001}, 
$n=1 - 6$ \cite {Xu2004}, 
$n = 12 - 29$ \cite {Yoo2006}, 
$n = 40 - 44$ \cite{Qin2009}
and, for smaller clusters, 
as a function of charge \cite{Li2001,Xu2004,King2002,King2003,King2006,King2007}.  
Techniques used typically include DFT.  
As the number $n$ increases, 
studies focus on interesting ``motifs,'' e.g., plate-like \cite{Qin2009}
and structures that are building blocks for larger clusters \cite {Zhao2008}, 
and because of large size, are searched for with genetic algorithms.  
Experimental gas-phase cluster studies show an evolution from prolate to more spherical geometries occurring at 
$n \approx 40$ \cite{Hunter1994}.  
Au-Ge nanoparticles have been recently 
fabricated using laser ablation in water \cite {Musaev2012},
finding clusters with a chainlike morphology.  
Recently, Li {\it et al.} \cite{Li2009,Li2013}  
have studied Au-doped Ge$_n$ clusters with $n=1 - 13$,
confirming that small gas phase Au-Ge clusters are amorphous,  
as observed experimentally \cite{Kingcade1979}.
More recently, theorists have focused on endohedral Ge clusters 
encapsulating a metal ion \cite{Korber2009,Tai2011}. 

Interestingly, 
experimentalists have formed anionic clusters AuGe$_{18}$ \cite{Schenk2007}, 
$\mathrm{Au_3Ge_{18}^{5-}}$ \cite{Spiekermann2007a}, and 
$\mathrm{Au_3Ge_{45}^{9-}}$ \cite{Spiekermann2007}; 
all have a building block  $\mathrm{Ge_9^{q-}}$.  
As studied in Zintl ion chemistry, these building blocks  are 
found in solids such as $\mathrm{Cs_4Ge_9}$ \cite{Sevov2002}.  
These same building blocks  provide a method of synthesizing crystalline 
and micro-crystalline Ge solids \cite{Armatas2006,Sun2006} 
and the clathrate-II phase $\mathrm{Ge_{136}}$ \cite{Guloy2006}.  
{\it allo-\/}Ge\cite{Kiefer2011} is another Zintl-ion example, 
formed by removing Li from a Zintl compound Li$_7$Ge$_{12}$.
Ge Zintl ions are stabilized by delocalized electrons, 
follow Wade's Rules 
(also known as Wade-Mingos' Rules) \cite{Wade1971,Mingos1972,Welch2013},  
and reveal bonding environments 
beyond the $\mathrm{sp^3}$ bonds of bulk Ge structures \cite{Zintl1939}.
This is the starting point of our study.  

Wade studied the element Boron, 
trying to understand its molecular structure as a function of charge state 
and observed molecular clusters that formed in polyhedral \\shapes, 
either as a complete polyhedron or one missing one or more vertices.
The number of vertices was counted by Wade as $n-1$, 
with $n+1$ being related to the number of skeletal electron pairs (SEPs).
Closed  polyhedra were referred to as {\it closo}.  
One missing a single vertex was termed ``nestlike,'' {\it nido},
and one missing two vertices was ``spider-like,'' {\it arachno}.  
Often, Wade's rules are rewritten, 
with the number of vertices now becoming $n$ 
so that {\it closo}, {\it nido}, and {\it arachno} clusters have
$(n+1)$, $(n+2)$, and $(n+3)$ SEPs, respectively.  
The deltahedral cages with 5 to 12 vertices 
are termed ``trigonal bipyramid'' (5), ``octahedron'' (6),
``pentagonal bipyramid'' (7), ``dodecahedron'' 
(8), ``tricapped trigonal prism'' (9), ``bicapped square-antiprism'' 
(10), ``octadecahedron'' (11), and ``icosahedron'' (12).
Note that B$_6$H$_6^{2-}$ is structurally and orbitally similar to Ge$_6^{2-}$, 
and should have octahedral, $O_h$ symmetry.  
Counting available electrons in B$_6$H$_6^{2-}$, 
each B provides 3 and each H provides 1, 
so the total is $26= 6 \times 4+2$.
The six covalent B-H bonds use up 12 electrons, 
leaving 14 electrons for 12 B-B bonds.  
Traditional two-center two-electron ($2c2e$) bonds 
cannot stabilize this structure.  
The electrons are thus delocalized.  
Counting the  skeletal electron pairs, $14/2=7$,
so a {\it closo} octahedron can be formed with Wade's Rules.

In this paper, 
we focus our work on $\mathrm{Ge_m^{q-}}$ clusters, 
where $\mathrm{m = 6, 9, 12}$, 
with emphasis on the less studied \GeM{6}\\cage geometry.  
We examine the stability of these cages 
when a bridge of one to three Au atoms is used to connect 
two nominally identical $\mathrm{Ge_m^{q-}}$ clusters.  
The SIESTA techniques used
include Conjugate Gradient (CG) minimization, 
study of the electronic properties using densities of states and 
an electronic partitioning method, visualization of the molecular orbitals,
Crystal Overlap Hamiltonian Population (COHP)~\cite{Dronskowski1993},
and molecular dynamics (MD).

\section{Methods}
\label{sec:methods}

    We have carried out electronic structure calculations
    with the fully \textit{ab initio} 
    DFT code SIESTA \cite {Sanchez-Portal1997,Soler2002} 
    which uses Troullier-Martins
    norm-conserving pseudopotentials \cite{Troullier1991} 
    in the Kleinman-Bylander form \cite{Kleinman1982}. 
    The Kohn-Sham \cite{Kohn1965} wave function 
    was expanded with basis sets of the 
    localized atomic orbitals (LCAO) 
    of the method by Sankey and Niklewski \cite{Sankey1989}. 
    The localization is a key advantage 
    in studying charged states in molecular or nanosized systems. 
    We use double-zeta polarized (DZP) orbitals 
    and the Generalized Gradient Approximation (GGA)
    of Perdew, Burke, and Ernzerhof \cite{Perdew1996} 
    for the exchange-correlation energy functional.

    The Ge basis set has cutoff radii for the 
    $4s$ orbital of 7.12, 1.92 Bohr; 
    $4p$ of 7.72, 4.92 Bohr;  
    and a polarization orbital 
    $4d$ with a cutoff of 7.79 Bohr \cite{Anglada2002}.  
    The Au basis set has cutoff radii of 
    $5d$: 7.20, 5.57 Bohr;
    $6s$: 6.50, 5.50 Bohr; 
    and a polarization orbital $6p$: 5.85 Bohr.
    Soft-confinement potentials 
    and an ionic core are used 
    to create basis orbitals \cite{siesta_manual}.  
    The Au pseudopotential includes the semicore $5d^{10}$ states
    \cite{Fernandez2004,Fernandez2006,Torres2005}.
    Both Au and Ge use a partial-core correction 
    to treat overlap between valence and core electrons. 

    The charge density is represented 
    on a real-space grid with an energy cutoff of 100 Ry. 
    A Monkhorst-Pack $k$-point mesh of 10x10x10 ensured convergence. 
    We used a Conjugate Gradient (CG)
    minimization with a maximum force tolerance of 0.04 eV/\AA~
    to reveal the local coordinate relaxations. 
    For our molecular dynamics (MD) runs, 
    we use the Nos\'{e} thermostat~\cite{Nose1984} 
    to simulate the canonical ensemble (NVT) 
    with a Nos\'{e} mass of $200~\mathrm{Ry~fs^2}$.
    We test the stability of our systems by subjecting them 
    to a temperature of 600K during 1000 steps, 
    each of 2 fs, as suggested by Ref.~\cite{Fernandez2006}.
       
     The COHP uses DFT matrix elements to
    separate electronic bonding states 
    from antibonding states 
    in the DOS.  
    In both the DOS and COHP curves 
    we use a small smearing value, $s\approx0.1$,
    and a high number of sampling points, $n\approx500$,  
    to create smooth electronic structure curves.  
    We treat neighbor interactions within 0.4~\AA~ 
    of the maximal considered ``bond'' lengths \cite{fn}.
       In examining the electronic states, 
    we focus on the hybridization of the levels presented 
    as indicators of electron delocalization. 
 
\section{Results}
\label{sec:results}
\subsection{Ge$_6$ and Ge$_9$ Clusters}
\label{sec:ge6_9}

Ge is an example of a main-group element 
that participates in delocalized cage bonding.   
\GeM{6}~cages have been experimentally synthesized 
in gas phase \cite{Fassler1998} and solids \cite{Kircher1998,Renner2000}.
Richards \textit{et al.} isolated the octahedral \GeM{6}~cluster,  
stabilized by organic groups, with reductive coupling \cite{Richards2003}.
\GeM{6}~cages have been studied computationally \cite{Ogut1997,King2002} 
with a focus on the effects of ionic charge on cluster geometry: 
neutral \GeM{6}~flattens to a $C_{2v}$ symmetry 
while ionic Ge$_6^{2-}$ has octahedral $O_h$ symmetry,
as predicted by Wade's Rules.
The neutral \GeM{6}~geometry 
was also observed by Zhao \textit {et al.} \cite{Zhao2008} 
in a computational study which focused primarily on fragmentation behavior 
in $\mathrm{Ge_n}$ clusters, 
and indicates \GeM{6} as a common fragment of larger clusters.           
Tantalizingly, two types of chains of \GeM{9} have been studied experimentally, 
(-Ge$_9^{2-}$-)$_\infty$ \cite{Ugrinov2005,Nienhaus2006}
and a \\trimer having the form [Ge$_9$=Ge$_9$=Ge$_9$]$^{6-}$ \cite{Ugrinov2002}.
Theorists have analyzed 
the localized bonds in these chains \cite{Pancharatna2006}
and examined nanoclusters based on Ge$_9$ clusters \cite{Karttunen2010}.

Using CG minimization in SIESTA, 
we initialize the geometries of 
$\mathrm{Ge_6}$ and $\mathrm{[Ge_6]^{2-}}$ clusters 
as octahedra with atomic coordinates 
$\pm(a,0,0)$,  $\pm(0,a,0)$,  $\pm(0,0,a)$  using $a \approx 2.12$~\AA, 
or initial Ge-Ge separations of 3.0~\AA. 
Our relaxed clusters 
have the same geometries as King {\textit{et al.} \cite{King2002}
Calculating a minimum and maximum relaxed Ge-Ge distance in $\mathrm{Ge_6}$, 
we find respectively 2.53~\AA~ and 2.85~\AA, 
compared with previous calculations of 
2.58~\AA~ and 2.81~\AA~ \cite{King2002} 
and 2.47~\AA~ and 2.85~\AA~ \cite{Ogut1997}.  
Our calculated bond length, 2.68~\AA, for octahedral $\mathrm{[Ge_6]^{2-}}$  
compares favorably with that of 2.69~\AA~ from  Ref.~\cite{King2002}, 
and has been measured to be 2.63~\AA~\cite{Richards2003} 
in a ligand stabilized system.

Our calculated total density of states (DOS) 
and partial density of states (pDOS)  
of (a) $\mathrm{Ge_{6}}$ and (b) $\mathrm{Ge_{6}^{2-}}$ 
are shown in Fig.~\ref{fig:dosge6}.   
The pDOS shows orbitals dominated 
by $4s$ character below $E-E_F = -2.0~\mathrm{eV}$ and 
by $4p$ in the range $-2.0~\mathrm{eV} < E-E_F < 0.0~\mathrm{eV}$.  
In the neutral $\mathrm{Ge_6}$ cluster, 
we see a stronger hybridization of atomic states, 
as shown in Fig.~\ref{fig:dosge6}(a) when compared 
with (b) $\mathrm{[Ge_6]^{2-}}$.
This is further confirmed by the MOs, shown in Fig.~\ref{fig:mo-ge6}
along with the corresponding energy levels (within $\pm0.1$ eV).
Our calculated MO energy levels for $\mathrm{Ge_6}$ and $\mathrm{[Ge_6]^{2-}}$  
compare favorably to the results found by King \textit {et al.} \cite{King2002}.   
The overall symmetries and degeneracies match Table~5 in Ref.~\cite{King2002}, 
however, due to different basis sets,  we find a consistent energy shift of
approximately 2-3 eV for  $\mathrm{[Ge_6]^{2-}}$  
and 5.5-6.5 eV for $\mathrm{Ge_6}$.       
The two additional electrons in $\mathrm{[Ge_6]^{2-}}$  
shift the Fermi energy, $E_F$, upward 
and allow a thirteenth orbital to be occupied.  
This presumably drives the symmetry change 
to $O_h$, with $p$ electrons now able to reside in a cubic environment.
    
We characterize cluster symmetry 
with a measure of deviation, $\sigma_{d_G}$,
of the cluster from a perfect octahedron. 
We average over the twelve skeletal 
Ge-Ge distances of a single cluster, 
calculate the standard deviation $\sigma_{d_G}$ 
of this value, 
and report $\bar{d}_G \pm \sigma_{d_G}$.
A perfect octahedral cluster 
is composed of twelve identical bonds with $\sigma_{d_G} = 0.0$~\AA.  
Some deformation from $O_h$ symmetry 
may occur even while $\sigma_{d_G}$ remains zero; for instance we observe 
$\mathrm{Ge_{6}^{2-}}$ has nearly identical bondlengths 
of $\bar{d}_G = 2.68$~\AA~such that $\sigma_{d_G} = 0.0$~\AA~
yet it is slightly flattened.  
The neutral $\mathrm{Ge_{6}}$ cluster 
with $C_{2v}$ symmetry yields  $\bar{d}_G = 2.68 \pm 0.20 $~\AA, 
the agreement of the average bondlength 
is coincidental since the neutral cluster is composed 
of eight short bonds $d \approx 2.5$~\AA~ 
and four long bonds $d \approx 2.8$~\AA. 
    
\begin{figure}
\resizebox{0.5\textwidth}{!}{%
  \includegraphics{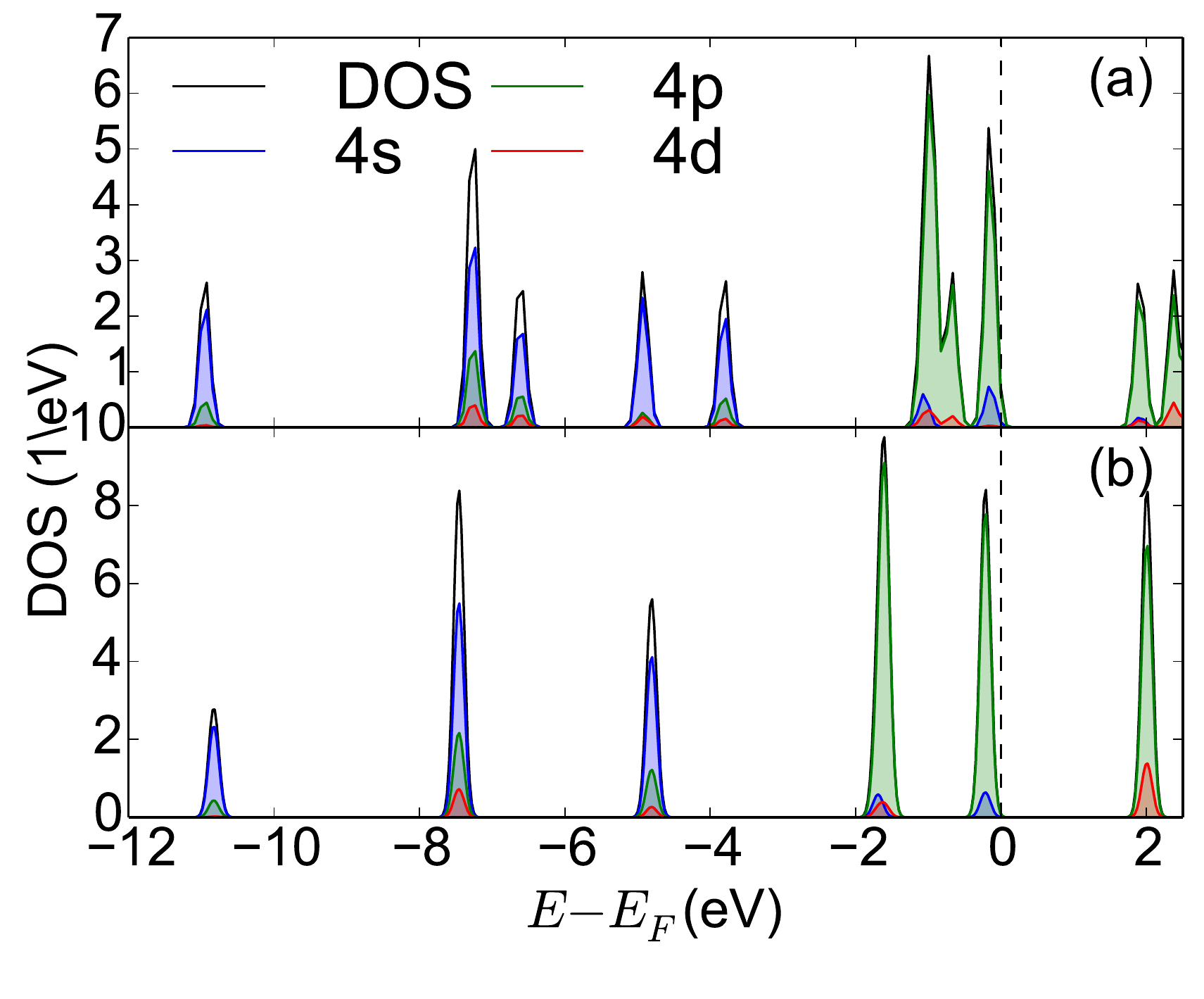}
}
\caption{Calculated densities of states (DOS) versus energy (eV) for 
         (a) $\mathrm{Ge_6}$ and (b) $\mathrm{[Ge_6]^{2-}}$ 
         where the total DOS is black, 
         the $4s$ orbital is blue, 
         $4p$ is green, 
         and $4d$ is red.  
} 
\label{fig:dosge6}       
\end{figure}
%
\begin{figure}
\resizebox{0.5\textwidth}{!}{%
  \includegraphics{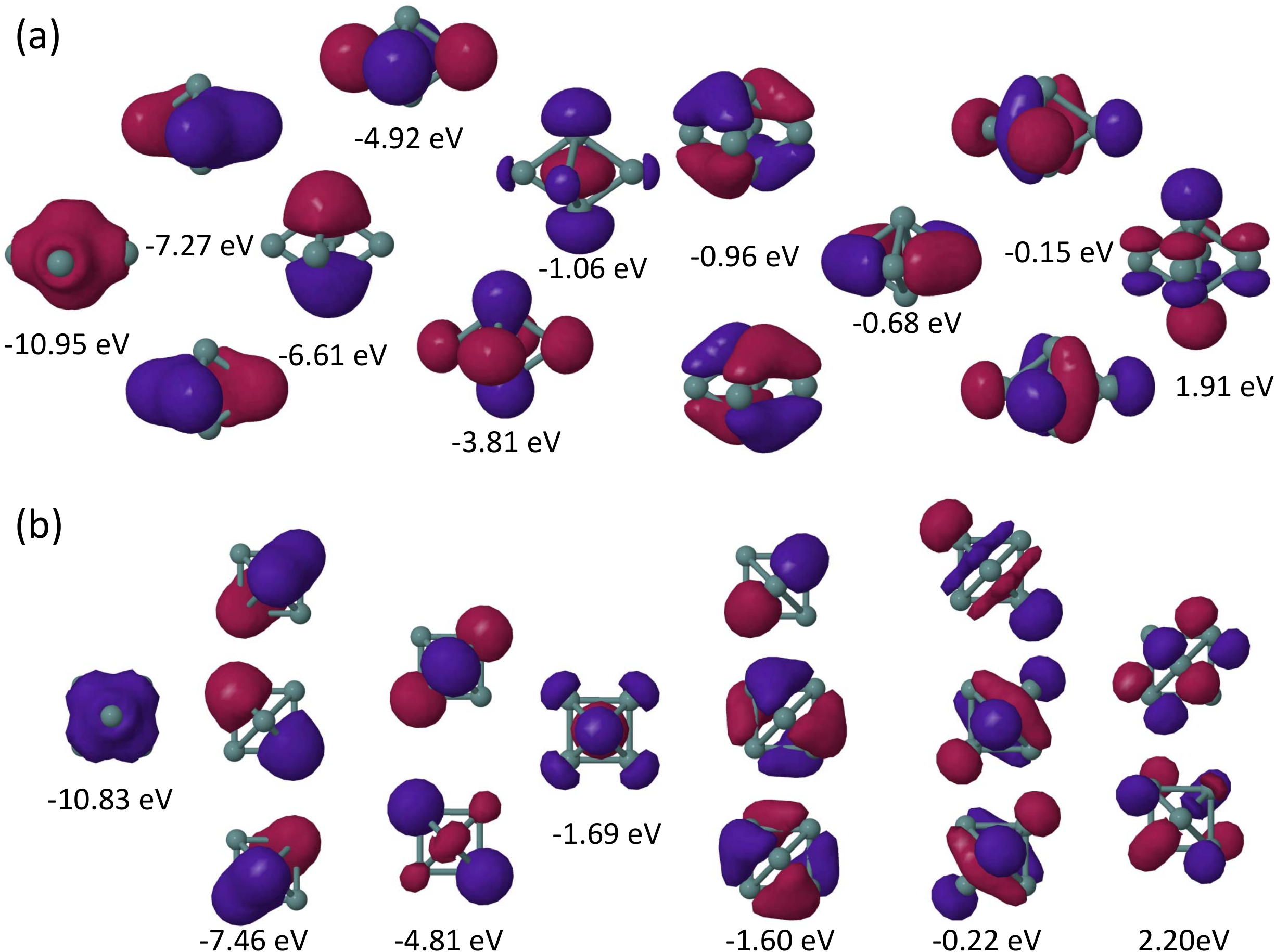}
}
\caption{%
Molecular orbitals of (a) $\mathrm{Ge_6}$ and (b)  $\mathrm{Ge_6}^{2-}$.
} 
\label{fig:mo-ge6}       
\end{figure}

For  $\mathrm{Ge_{9}}$, 
we initialize the CG relaxation using previously reported geometries 
and observe very similar clusters to those previously calculated 
in Refs.~\cite{Zhao2008,King2007,Bulusu2005,King2003} 
both in geometry and electronic structure.  
Consistent with Wade's rules, 
we observe the symmetry of Ge$_9^{q-}$ increases 
to $D_{3v}$ with $q=2$ and $C_{4h}$ with $q=4$ 
as in Ref.~\cite{King2003}.  
We see a qualitative agreement in 
the MOs, with the same symmetries 
and approximate energies levels as previously reported.  

Binding energies $E_B$ 
and Fermi gaps $E_{\mathrm{gap}}$  are shown in Table~\ref{tab:ge6-9-12}
for the  \GeM{m}~clusters.
$E_B$ is calculated through out this paper as:  
\begin{equation}
     \label{eq:1}
     E_{B} = \frac{E_{T} - (N_AE_A + N_GE_G)}{N_A+N_G}
   \end{equation}
where $N_A$ and $N_G$ are the total number of Au and Ge atoms, 
and $E_{T}$,  $E_A$, and $E_G$ are 
the SIESTA calculated total energies 
for the cluster and isolated atoms, respectively. 
Our calculation shows systematically higher $E_B$, 
than previously reported for pure Ge clusters.  
The systematic difference in $E_B$ 
is consistent with cohesive energies 
calculated with SIESTA in Ref.~\cite{Soler2002} 
for bulk Si and elsewhere. 
While our values of $E_{\mathrm{gap}}$ differ, 
given slight deformations in the geometries, 
and the known underestimates 
of $E_{\mathrm{gap}}$ within GGA calculations, 
we consider our values 
reasonable comparisons to previous calculations.   

\begin{table}
\caption{Calculated $E_B$ and $E_{\mathrm{gap}}$ (eV) 
  of CG-relaxed clusters 
  $\mathrm{Ge_{6}}$, $\mathrm{Ge_{9}}$, 
  and $\mathrm{Ge_{12}}$ (Z$_0$) 
  from this work and those of Ref.~\cite{Zhao2008} (in parentheses).  }
\label{tab:ge6-9-12}      
\begin{center}
\begin{tabular}{lllllll}
\hline\noalign{\smallskip}
        & \multicolumn{2}{c}{$\mathrm{Ge_{6}}$} & 
          \multicolumn{2}{c}{$\mathrm{Ge_{9}}$} & 
          \multicolumn{2}{c}{$\mathrm{Ge_{12}}$ ($Z_0$)} \\
\noalign{\smallskip}\hline\noalign{\smallskip}
$E_B$    & 4.425 & (3.092) & 4.589 & (3.215)  & 4.650 & (3.270)   \\
$E_{\mathrm{gap}}$ & 2.05 & (1.992) & 1.24 & (1.676) & 1.59 & (2.003)  \\
\noalign{\smallskip}\hline
\end{tabular}
\end{center}
\end{table}
\subsection{Ge$_{12}$ Clusters}
\label{sec:Ge-clusters}

Zhao \textit {et al.} \cite{Zhao2008} found 
the lowest energy isomer of the $\mathrm{Ge_{12}}$ cluster to be 
a tetracapped cube with $C_{2v}$ symmetry, Fig.~\ref{fig:cgmdge12}(a). 
King \textit {et al.} \cite{King2007} 
have also studied this cluster and note that it deviates from Wade's Rules,
it is \textit{not} an icosahedron, 
which would require vertex degrees higher than 4.
We also find the same structure found by these authors, 
with qualitative agreement in 
the MOs, and the same symmetries and approximate energies levels.  

Here we examine the interesting possibility that 
isomers of $\mathrm{Ge_{12}}$ may form by joining two $\mathrm{Ge_6}$ cages 
together with one or two intercluster bonds. 
While these will not result in groundstate \GeM{12} clusters, they are analogous
to the \GeM{9} chains found experimentally.
We initialize each $\mathrm{Ge_6}$ cluster as an octahedron,  
again with an initial Ge-Ge separation of 3.0~\AA, 
and change their relative orientation.

Three initial configurations of
$\mathrm{Ge_6}$-$\mathrm{Ge_6}$ clusters were considered 
and were named I$_n$, II$_n$, etc. 
for their relative energy (after CG relaxation) 
in the order of highest to lowest binding energy
(the notation is general, allowing for clusters that include $n$ Au atoms).
Here we describe the input geometries of the clusters, 
but their structures can be easily understood 
from Fig.~\ref{fig:cgmdge12}(b-d) 
since the CG relaxed output geometries 
differ only slightly from their inputs.  
Using two local Cartesian coordinate systems 
$(x,y,z)$ and $(x^\prime,y^\prime,z^\prime)$ 
for each cluster, 
in configuration I$_0$ [Fig.~\ref{fig:cgmdge12} (b)], 
the axes $(x,y,z)$ and $(x^\prime,y^\prime,z^\prime)$  
differ by a linear translation so that all axes  
point respectively in the same directions and 
the atoms in the two equatorial planes $xy$ and ${x^\prime}{y^\prime}$ 
form two intercluster Ge-Ge links 
pointing locally along directions 
$\hat{x}+\hat{y}$ and $-\hat{x}^\prime-\hat{y}^\prime$.
In configuration II$_0$ [Fig.~\ref{fig:cgmdge12} (c)], 
the two sets of axes again point in the same directions 
and a single intercluster link forms in the local directions 
$\hat{x}$  and $-\hat{x}^\prime$.  
In configuration III$_0$ [Fig.~\ref{fig:cgmdge12} (d)], 
the two octahedra are aligned so that 
two triangular faces of opposing octahedra are in parallel planes 
and three intercluster links form a prism.  

\begin{figure}
\resizebox{0.5\textwidth}{!}{%
  \includegraphics{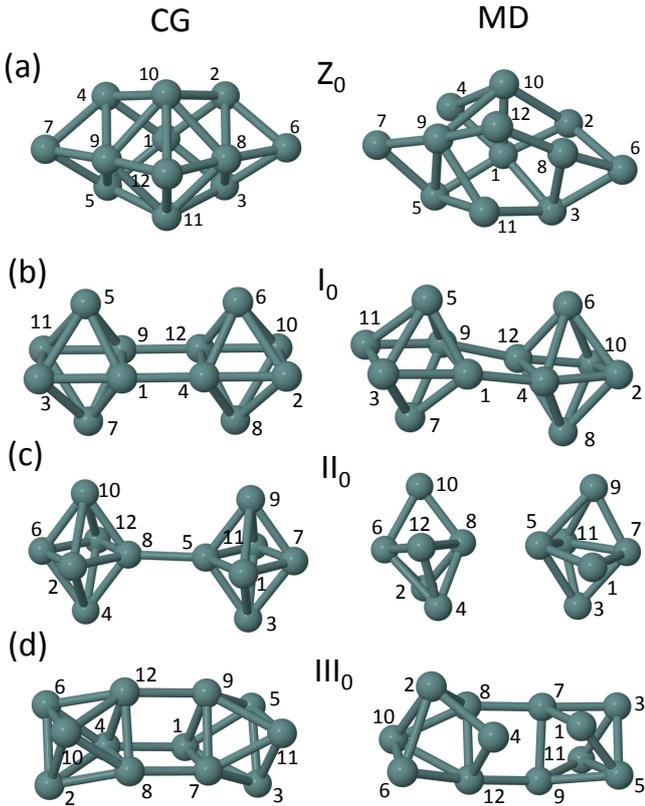}
}
\caption{$\mathrm{Ge_{12}}$ clusters 
         (a) Z$_0$; (b) I$_0$; (c) II$_0$; and (d) III$_0$. 
         The left column shows 
         relaxed geometries from 
         CG minimization and the
         right shows the clusters after a MD run. 
         The structures are all rotated slightly 
         to show the geometry of the interconnections\cite{fn}.
         }
\label{fig:cgmdge12}   
\end{figure}

In Fig.~\ref{fig:cgmdge12} we show 
the CG-relaxed (first column) and 
MD ``temperature shaken'' (second column) geometries of each isomer.
MD simulations are started with the CG-relaxed structures shown in Fig.~\ref{fig:cgmdge12} 
and are run at 600K for 1000 timesteps of 2 fs each \cite{CRC}.  The numbers shown in the figure
show the location of each atom through the MD run. 
The primary differences between 
CG-relaxed and post-MD structures are small deformations 
that lead to an overall symmetry loss, as summarized in Table~\ref{tab:cgge12} and Table~\ref{tab:mdge12}.
Since Ge-Ge bonds elongate considerably 
in the cage environment supported with delocalized bonding, 
we do not attempt to distinguish bonds based on their distances \cite{fn}.

After CG relaxation, the values of  $E_B$ of all isomers 
are close to that of the ground state, which we labelled as  Z$_0$ 
since it is outside our schema of coupled $\mathrm{Ge_6}$ clusters.
We note first that 
the CG-relaxed Z$_0$ tetracapped cube has a relatively compact structure 
including a short Ge-Ge bond of 1.90~\AA, and no octahedral structure
(and thus we cannot measure $\bar{d}_G\pm\sigma_{d_G}$ or $d_i$ in Table~\ref{tab:cgge12}).
It retains much of its symmetry after a MD run.
The I$_0$, II$_0$, and III$_0$ clusters 
built from Wade octahedral $\mathrm{Ge_{6}}$ cages 
remain similar to their input geometries and 
including similar intercluster bonds, or interconnections. 
I$_0$ maintains two intercluster bonds $d_i$ of 2.56~\AA.  
II$_0$ has the most overall deformation with $\sigma = 0.182$~\AA~ 
and maintains one interconnection of length 2.68~\AA.  
Finally III$_0$ deforms so that 
it has one strong intercluster bond of $d_i$ = 2.64~\AA~ 
and two elongated interconnections of $d_i$ = 2.80~\AA~ 
which are much longer than a typical bulk Ge-Ge bond, 
but not unusual within a nanocluster or cage geometry.  
II$_0$ and III$_0$ are nearly degenerate energetically,
despite their different geometries.

After MD runs, 
I$_0$ retains two intercluster bonds and 
measurably octahedral $\mathrm{Ge_6}$ cages 
as shown in Fig.~\ref{fig:cgmdge12}(b).  
II$_0$ breaks its intercluster bond but its 
$\mathrm{Ge_6}$ clusters remain cage-shaped.  
With structure III$_0$, an intercluster bond breaks, 
and the relative tilting of the two cages has changed.  
In all cases, the deviation from octahedrality $\sigma_{d_G}$  
is larger in the post-MD structures, 
the overall symmetry of the cluster is reduced, 
and the average skeletal Ge-Ge bond length has increased, 
as quantified in Table~\ref{tab:mdge12}, 
which also shows the post-MD bondlengths of 
$\mathrm{Ge_6}$ and $\mathrm{Ge_6^{2-}}$ 
for comparison.

%
\begin{table}
\caption{Average skeletal distance $\bar{d}_G$, 
  deviation from octahedrality $\sigma_{d_G}$,
  range of bond lengths  $\sigma_{d_G}$, 
  intercluster distances $d_i$ (all distances in~\AA), 
  and binding energy per atom $E_B$ (eV) 
  calculated for the CG-relaxed clusters $\mathrm{Ge_{12}}$.  
}
\label{tab:cgge12}  
\begin{center}
\begin{tabular}{llllll}
\hline\noalign{\smallskip}
$\mathrm{Ge_{12}}$ &  $\bar{d}_G$ &   $\sigma_{d_G}$ & $d_G$ & $d_i$ & $E_B$\\
\noalign{\smallskip}\hline\noalign{\smallskip}
Z$_0$   & ------  & ------ & 1.90-2.80 & ------ & 4.650 \\
I$_0$   & 2.667 & 0.104 & 2.55-2.80 & 2.56 & 4.527 \\
II$_0$  & 2.683 & 0.182 & 2.55-2.93 & 2.68 & 4.475 \\
III$_0$ & 2.667 & 0.064 & 2.57-2.77 & 2.64-2.80 & 4.474 \\ 
\noalign{\smallskip}\hline
\end{tabular}
\end{center}
\end{table}
%

\begin{table}
\caption{Calculated average skeletal distances
and deviations from octahedrality (\AA) of 
the post-MD clusters $\mathrm{Ge_{6}}$, 
$\mathrm{Ge_{6}^{2-}}$, and $\mathrm{Ge_{12}}$.  }
\label{tab:mdge12}       
\begin{center}
\begin{tabular}{llllll}
\hline\noalign{\smallskip}
              & ${(\bar{d}_G)}_1$ & ${(\sigma_{d_G})}_1$ & $d_G$  \\  
\noalign{\smallskip}\hline\noalign{\smallskip}
\GeM{6}       & 2.72 & 0.27 & 2.48-3.36 \\
\GeM{6}$^{2-}$ & 2.69 & 0.03 & 2.63-2.72 \\
\hline\noalign{\smallskip}
$\mathrm{Ge_{12}}$  & ${(\bar{d}_G)}_1$ & ${(\sigma_{d_G})}_1$ & ${(\bar{d}_G)}_2$ & ${(\sigma_{d_G})}_2$ & $d_G$  \\  
\noalign{\smallskip}\hline\noalign{\smallskip}
Z$_0$   & ----- & ----- & ----- & ----- & 2.34-3.14 \\
I$_0$   & 2.721 & 0.139 & 2.757 & 0.173 & 2.49-3.00 \\
II$_0$  & 2.710 & 0.265 & 2.773 & 0.360 & 2.43-3.48 \\
III$_0$ & 2.767 & 0.292 & 2.732 & 0.278 & 2.51-3.35 \\
\noalign{\smallskip}\hline
\end{tabular}
\end{center}
\end{table}

Now we use the DOS (Fig.~\ref{fig:dos_ge12}) 
to examine the stability of 
CG-relaxed $\mathrm{Ge_{12}}$ structures.  
In Fig.~\ref{fig:dos_ge12}(a) 
we present the DOS of Z$_0$ where 
the $4s$ orbitals at $ -6.0 < E-E_F < -4.0$~eV 
remain well separated with a similar magnitude 
as neutral $\mathrm{Ge_{6}}$.  
The $4p$ states range over $-4.0 < E-E_F < 0.0$~eV, 
again of similar magnitude to $\mathrm{Ge_{6}}$, 
but with a much broader range, 
indicating the many possible orientations of $p$ orbitals 
in a tetracapped cube compared with an octahedron.  
The placement of the Fermi level at the center 
of the energy gap indicates this is a stable, 
insulating cluster.  

\begin{figure}
\resizebox{0.5\textwidth}{!}{%
  \includegraphics{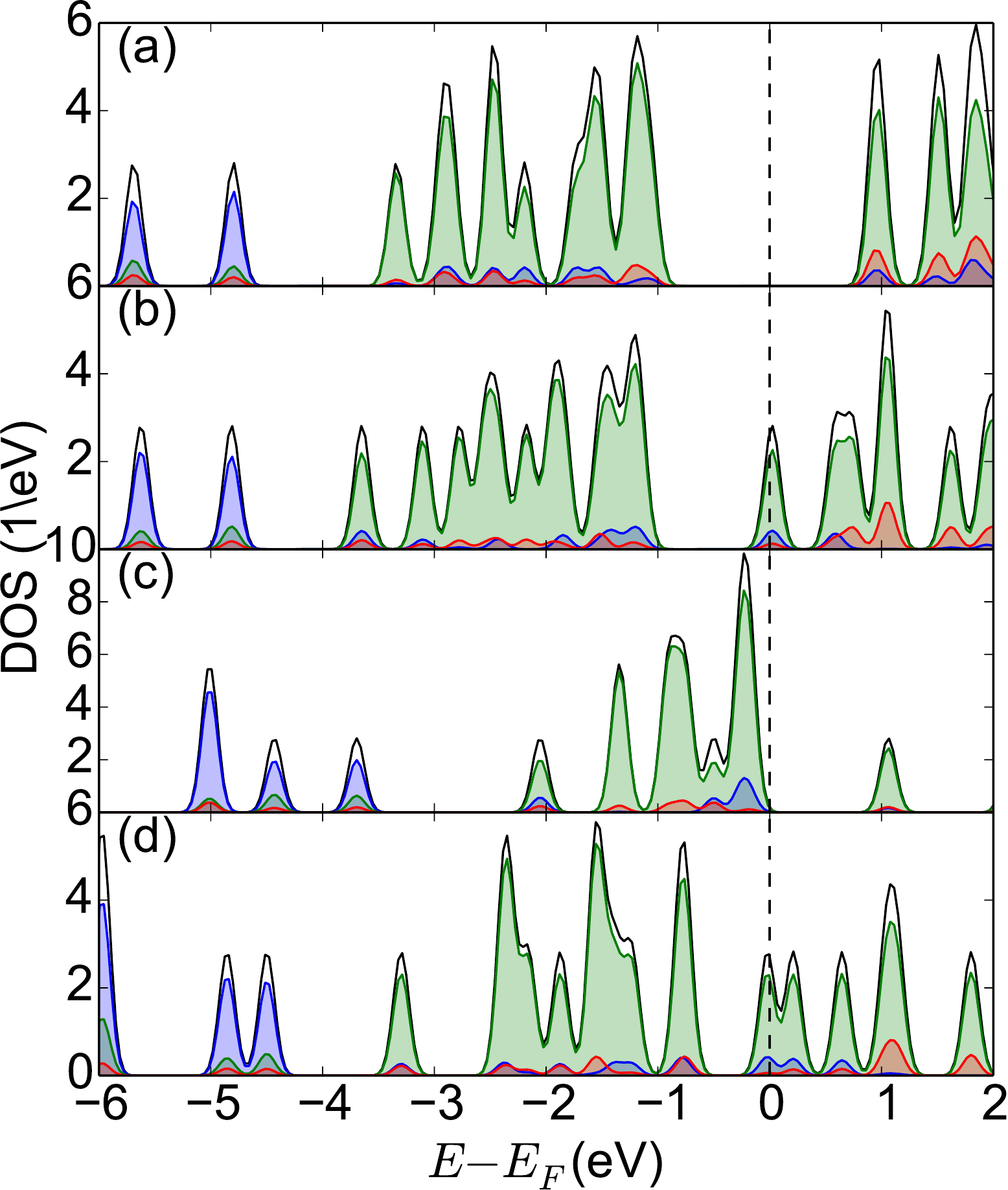}
}
\caption{DOS of CG-relaxed $\mathrm{Ge_{12}}$ isomers: 
  (a) Z$_0$, (b) I$_0$, (c) II$_0$, (d) III$_0$ 
  (see Fig.~\ref{fig:dosge6} for color key).  
}
\label{fig:dos_ge12}
\end{figure} 

In Fig.~\ref{fig:dos_ge12}(b), for I$_0$,
we see a similar set of $4s$ and $4p$ states below $E_F$, with additional states at $E_F$ itself.   
These are likely due to the overlap of the two intercluster Ge-Ge bonds.  
The placement of E$_F$ indicates these states are metallic.
Conversely in Fig.~\ref{fig:dos_ge12}(c), 
II$_0$ is similar to that of $\mathrm{Ge_6}$ in 
Fig.~\ref{fig:dosge6}(a).  While the two highest $4s$ orbitals 
remain of similar magnitude to those of the other $\mathrm{Ge_{12}}$ clusters,  
like $\mathrm{Ge_6}$ they are located at $-4.5 < E-E_F < -3.0$ eV.  
There is greater symmetry indicated in this cluster, with the $4p$ orbitals in the range $-2.5 < E-E_F < 0.0$ eV, 
and higher/fewer peaks than the other $\mathrm{Ge_{12}}$ clusters, 
again much like an individual $\mathrm{Ge_6}$ cluster.  
The II$_0$ cluster is insulating since E$_F$ is at the base of $E_{\mathrm{gap}}$ like neutral $\mathrm{Ge_6}$.  

As was shown with the MD simulation, II$_0$ is readily broken into 
two separated  $\mathrm{Ge_6}$ clusters, 
so it is unsurprising that the overall DOS is qualitatively similar to neutral $\mathrm{Ge_6}$.   
The additional peaks result from more available electronic states in \GeM{12} 
than $\mathrm{Ge_6}$, but we observe 
no states that may be clearly tied to the single intercluster bond.  

In Fig.~\ref{fig:dos_ge12}(d) we show that 
III$_0$ is metallic with many states in E$_{\mathrm{gap}}$.  
The DOS is similar to that of I$_0$, 
yet the $4s$ states at $-5.0 < E-E_F < -4.2~\mathrm{eV}$
are closer together, 
there are fewer peaks in the $4p$ orbitals, $-4.5 < E-E_F < -0.9~\mathrm{eV}$, 
and there is a double peak at $E_F$.  

We have found COHP electronic partitioning, shown in Figure~\ref{fig:cohpge12}, to be a useful additional
tool to determine bond stability (being computationally much less expensive than MD simulations).  
The groundstate $Z_0$, 
Fig.~\ref{fig:cohpge12}(a),   
introduces the typical criterion of a stable cluster: 
the states switch from negative bonding to 
positive antibonding MOs at the Fermi level. 
In structure I$_0$, Fig.~\ref{fig:cohpge12}(b), 
the anti-bonding states first appears 
at $E_F$ at the top of the band gap.  
Given the overall stability of this cluster (the two cluster-linking bonds remain stable during MD), 
this suggests the additional states introduced in the DOS by the intercluster bonds 
are antibonding, and serve to keep the cluster as two 
separate, linked structures.  
Structure II$_0$, Fig.~\ref{fig:cohpge12}(c),  
has antibonding states below $E_F$
at the base of the energy gap, which can be correlated with the 
dissociation into two $\mathrm{Ge_6}$ clusters in the post-MD structure.
Cluster III$_0$, Fig.~\ref{fig:cohpge12}(d),  appears metallic 
with the first antibonding state at $E_F$ (much like I$_0$) and a small energy gap.  
This is consistent with the post-MD rearrangements: the CG-relaxed structure of III$_0$
has three intercluster bonds and forms two during the MD run.  We  conclude that 
two intercluster Ge-Ge bonds provide a relatively stable 
connecting structure between $\mathrm{Ge_6}$ octahedral cages.  

\begin{figure}
\resizebox{0.5\textwidth}{!}{%
  \includegraphics{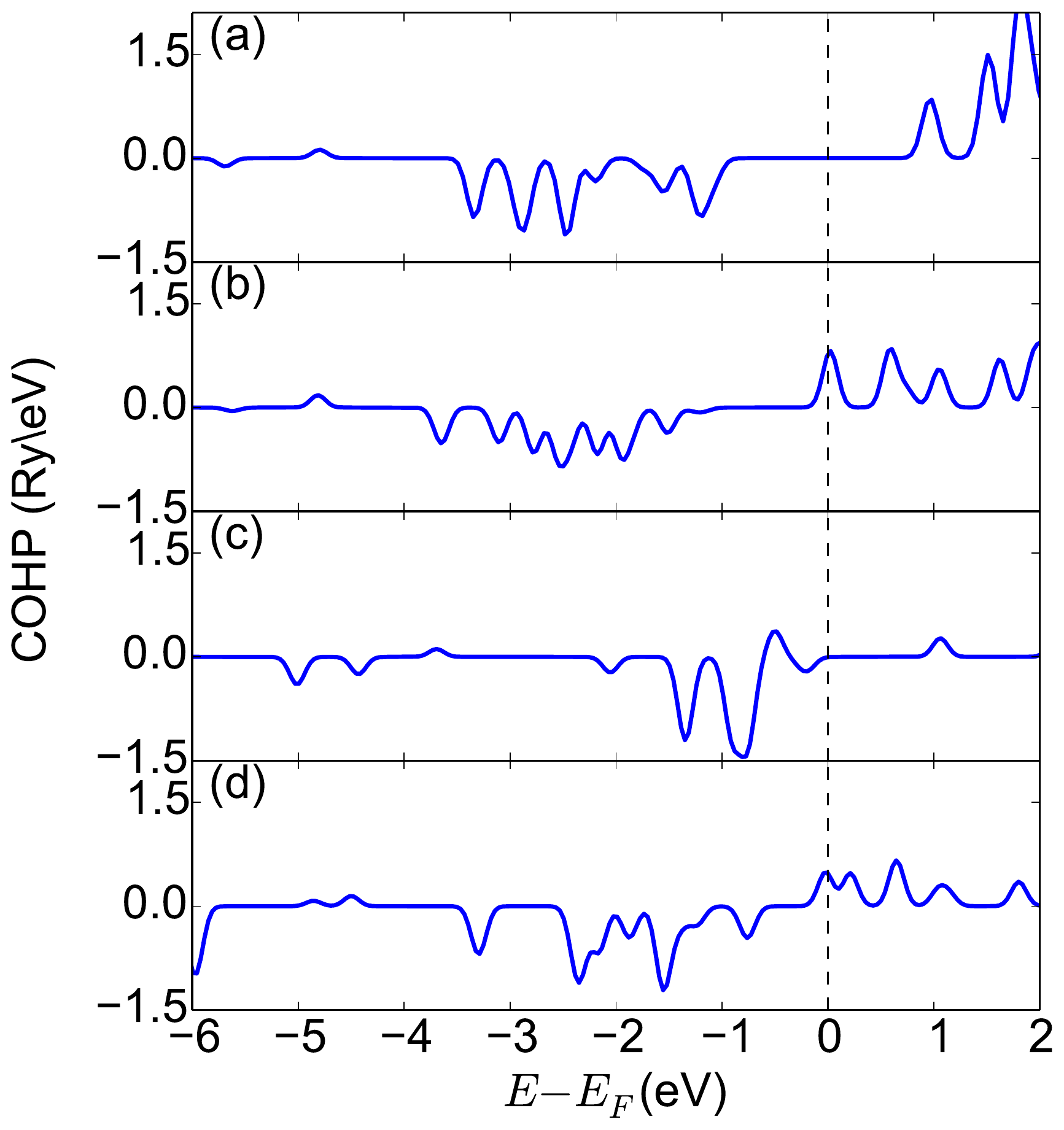}
}
\caption{COHP analysis of Ge-Ge interactions 
$\mathrm{Ge_{12}}$ after CG relaxation.  
Structures (a) Z$_0$, (b) I$_0$, (c) II$_0$, and (d) III$_0$. 
}
\label{fig:cohpge12}     
\end{figure}
%

\subsection{$\mathrm{Au_3Ge_{18}}$ Clusters}
\label{sec:au3ge18}

Next we study CG-relaxed structures 
for 
$\mathrm{[Au_3Ge_{18}]^{5-}}$ 
and its neutrally charged counterpart 
in vacuum to compare with the experimentally observed structure.  
As shown in Fig.~\ref{fig:struct_Au3Ge18}, 
this structure consists 
of two deltahedron-shaped $\mathrm{Ge_{9}}$ cages   
linked by three Au atoms.   
The three Au atoms bond to three
Ge atoms in each cage, 
forming a prism-like interconnection.  
Since this figure highlights the interconnection, 
we mention the asymmetry of the clusters: 
the apex of one $\mathrm{Ge_{9}}$ cage 
points along an apex of the interconnecting Au-triangle 
while the opposing cage's apex 
points in the opposite direction 
(see Ref.~\cite{Spiekermann2007}).
In comparing the neutral and charged 
CG-relaxed structures, 
we observe a very similar intercluster bridge, 
but the neutrally charged cluster has 
relatively elongated Ge-Ge bonds 
[Fig.~\ref{fig:struct_Au3Ge18}(a)], 
so that a complete \GeM{9}~deltahedron does not form.  
In Table~\ref{tab:au3ge18}, 
we summarize our results, 
comparing them with  
Spiekermann \textit{et al.} \cite{Spiekermann2007}, 
who reported bond lengths from 
experimental measurements and 
DFT calculations 
using a hybrid exchange and correlation (XC) functional.  
Experimental values for the Au-Au bond lengths, 
2.900 to 3.095~\AA, are shorter than calculated values.  
As in experiment, 
we observe that Ge-Ge bonds adjacent to Ge-Au bonds 
are considerably shorter than other Ge-Ge bonds. 
We also compare the three angles 
$\theta=\theta\mathrm{_{Ge-Au-Ge}}$ of the 
intercluster linking prism in the Table~\ref{tab:au3ge18}, 
finding them to be in reasonable agreement with experiment.   
We conclude that without the presence of  stabilizing ligands 
SIESTA has successfully 
reproduced the structures found experimentally. 

In Fig.~\ref{fig:struct_Au3Ge18} we also compare the post-MD structures 
of \AuGe{3}{18} and \AuGe{3}{18}$^{5-}$
to the CG structures.  
The neutral cluster forms two additional Au-Ge bonds, 
and the right hand \GeM{9} cage opens 
forming a larger cage including the three Au atoms.  
The \AuGe{3}{18}$^{5-}$ maintains the motif 
of two \GeM{9} cages separated to an Au triangle 
since it does not form new Au-Ge bonds.  
In both clusters the Ge-Ge cage bonds elongate 
and one Au-Au bond becomes widely separated.  

\begin{figure}
\resizebox{0.5\textwidth}{!}{%
  \includegraphics{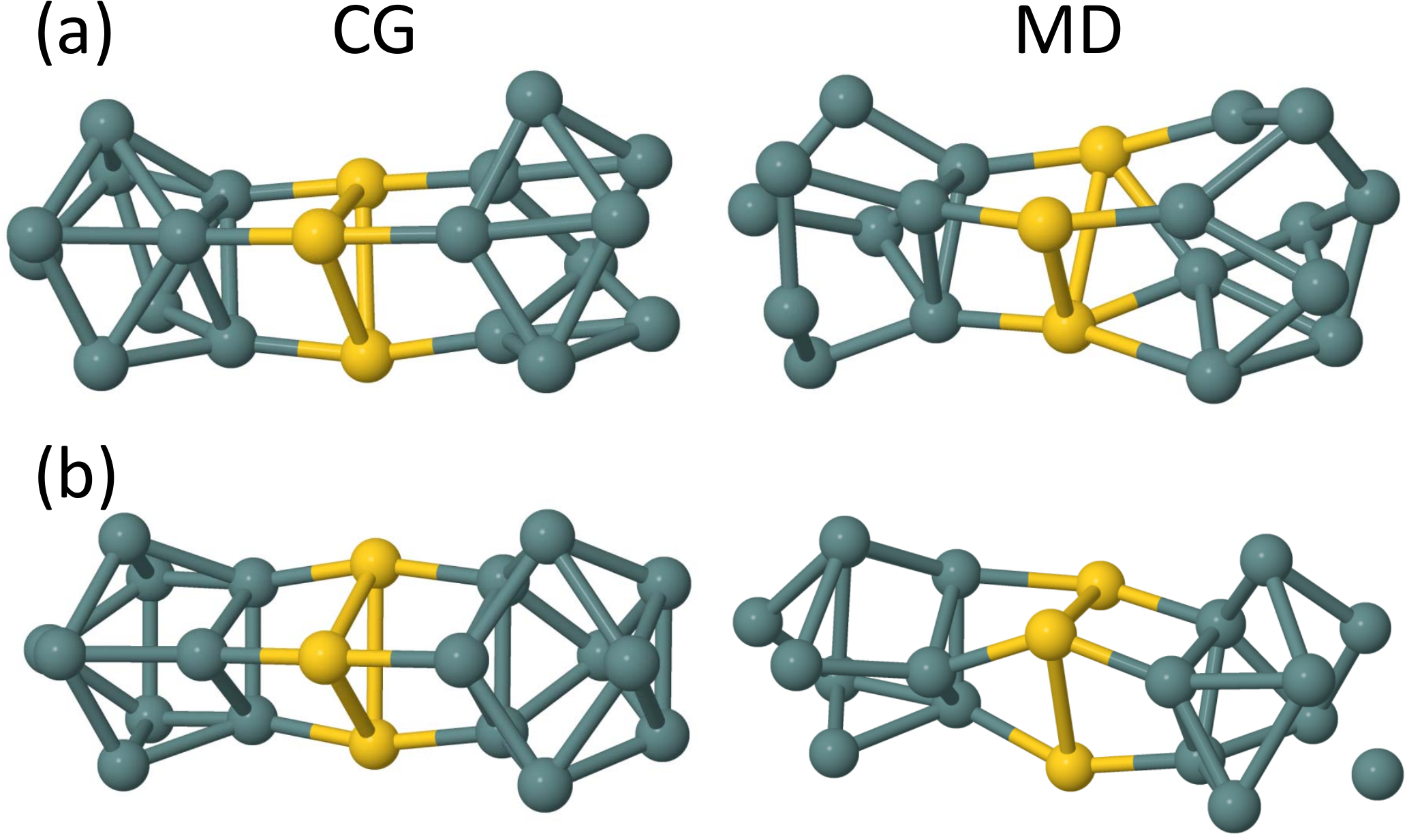}
}
\caption{$\mathrm{Au_3Ge_{18}}^{q-}$ clusters with charge 
         (a) $q = 0$ and (b) $q = 5$.  
         Au atoms are gold and Ge atoms are gray.
         }
\label{fig:struct_Au3Ge18}   
\end{figure}

We calculated the binding energy 
$E_B$ for the neutral cluster as 4.427 eV, 
which compares well other calculations in this work.
Our calculated $E_{\mathrm{gap}}$ 
is considerably smaller than 
reported by Ref.~\cite{Spiekermann2007}.  
Given the superiority of hybrid XC functionals 
for estimating optical gaps, 
this is unsurprising.  
However since our calculations show a 
stable cluster with a 
well-separated energy gap with the Fermi level at its base, 
which indicates overall cluster stability. 

In Fig.~\ref{fig:au3ge18_dos} we present  
(a) the total DOS and species pDOS,  
(b) the Ge pDOS of orbitals 
$4s$, $4p$, and $4d$, and 
(c) the Au pDOS of orbitals 
$5d$, $6s$, and $6p$ of the neutral cluster.  
In Fig.~\ref{fig:au3ge18_dos}(a) we see 
the strongest Au-Ge mixing appears just below $E_F$ 
where the Au $5d$ states mix with the Ge $4p$ states.
Below $E - E_F < -5.0~\mathrm{eV}$ the states are dominated by Ge $4s$ 
and between $-5.0 < E - E_F < -2.0$~eV 
they are dominated by Au $5d$ states.  
Here $E_F$ falls just below a larger gap, 
indicating a weakly conducting or insulating cluster.  
In Fig.~\ref{fig:au3ge18_dos}(b) we observe similar Ge orbitals 
to those in the pure $\mathrm{Ge_{12}}$ clusters 
with well separated $4s$ and $4p$ states.
One sees these $4p$ orbitals have considerable 
hybridization and peak at $E-E_F \approx 1.5$~eV, 
much like the neutral $\mathrm{Ge_6}$ cluster.
In Fig.~\ref{fig:au3ge18_dos}(c) 
one sees the $5d$ orbitals of Au in the region 
from $-4.5 < E-E_F < 0.5$~eV, 
which only strongly mix with the $6s$ orbital 
in the range $-1.5 < E-E_F < 0.5$~eV.  
The Fermi level is just below $E_{\mathrm{gap}}$, 
indicating a weakly metallic nature in this cluster, 
and the highest occupied MO is of mainly $4p$ character, 
as shown in Fig.~\ref{fig:au3ge18_HL}. 
 
The electronic structure of $\mathrm{Au_3Ge_{18}^{5-}}$ 
is quite similar to the neutral 
cluster, however $E_F$ sits at the top of $E_{\mathrm{gap}}$, 
which indicates it is unstable without external ligands.  
Indeed, as shown in Fig.~\ref{fig:struct_Au3Ge18}, 
we observed considerable rearrangement 
of both neutral and charged clusters under MD simulation, 
which restores the electronic structure to 
a stable configuration. 
Like its neutral counterpart, 
the pDOS of $\mathrm{Au_3Ge_{18}^{5-}}$ 
shows orbitals dominated by $4s$ character below $E-E_F = -2.0~\mathrm{eV}$ and 
by $4p$ in the range $-2.0~\mathrm{eV} < E-E_F < 0.0~\mathrm{eV}$.  
This is the 
characteristic orbital behavior 
in both pure Ge clusters and Au-Ge clusters 
which we observe throughout this study.  

In the COHP analysis of the neutral $\mathrm{Au_3Ge_{18}}$, 
Fig.~\ref{fig:au3ge18_COHP}, we note three features: 
in (a) the Au-Au interaction 
has strong antibonding states at $E_F$;
in (b) Au-Ge has a weak antibonding state at $E_F$
(this disappears in the post-MD structure); 
and in (c) Ge-Ge is bonding up to $E_F$, 
and antibonding above.  
At this level of theory, 
it appears Ge-Ge interactions are the strongest 
indicator of stability.  

\begin{figure}
\resizebox{0.5\textwidth}{!}{%
  \includegraphics{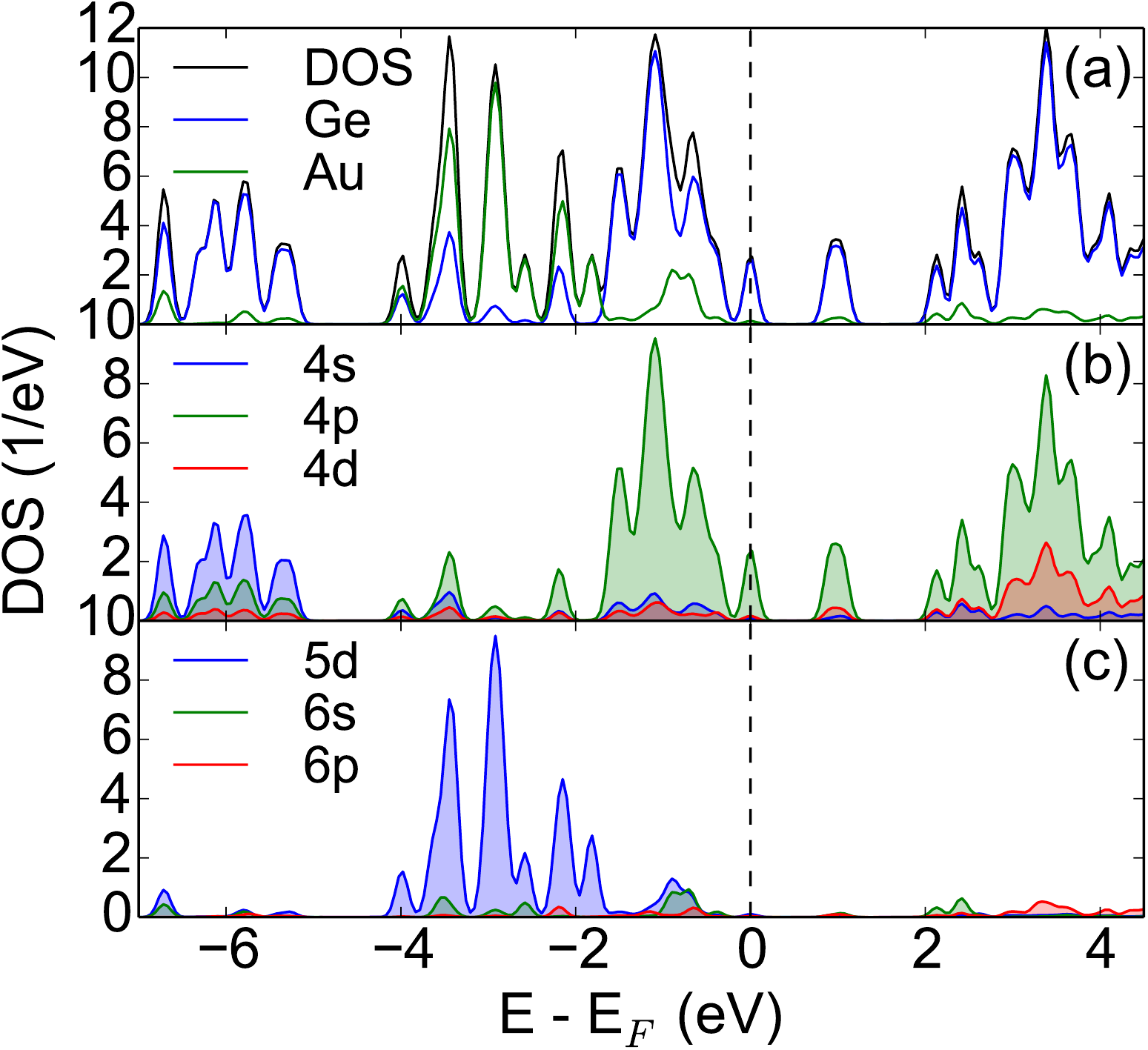}
}
\caption{The electronic structure of $\mathrm{Au_3Ge_{18}}$ (neutrally charged)
  (a) DOS (black), Ge pDOS (blue) and Au pDOS (green), 
  (b) Ge $4s$ (blue), $4p$ (green), and $4d$ (red), and 
  (c) Au $5d$ (blue), $6s$ (green), and $6p$ (red).  
}
\label{fig:au3ge18_dos}  
\end{figure}  
%

\begin{figure}
\resizebox{0.5\textwidth}{!}{%
  \includegraphics{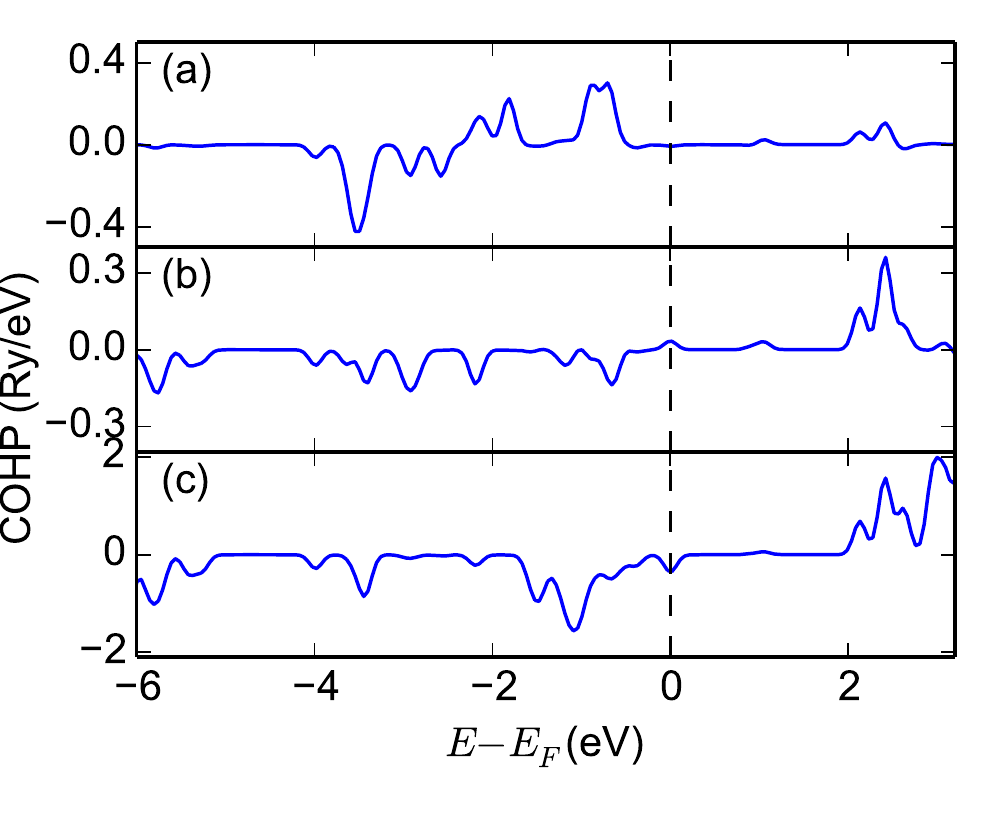}
}
\caption{COHP for $\mathrm{Au_3Ge_{18}}$
where for interactions 
(a) Au-Au, (b) Au-Ge, and (c) Ge-Ge.} 
\label{fig:au3ge18_COHP}       
\end{figure}
%

\begin{table}
\caption{Ranges of calculated bond lengths 
  (\AA) found for Ge, Au-Ge, and Au-Au bonds 
  for $\mathrm{[Au_3Ge_{18}]^{q-}}$.  
  Also shown, values for the angles 
  $\theta$ connecting Au to the two cages. }
\label{tab:au3ge18}
\begin{center}
\begin{tabular}{llllll}
\hline\noalign{\smallskip}
  $q$               & $d_G$ & $d_{AG}$ 
                         & $d_A$ & $\theta$
                         &  Ref.\\  
\noalign{\smallskip}\hline\noalign{\smallskip}
 0 & 2.5-2.85 & 2.46-2.48 
                         & 3.11-3.19 & 165$^{\circ}$-174$^{\circ}$ 
                         &  \\ 

       5- & 2.56-2.86 & 2.53 
                         & 3.03-3.34 & 165$^{\circ}$-172$^{\circ}$ 
                         &  \\
       5- & 2.55-2.88 & 2.45-2.46    
                         & 2.900-3.095  &  168$^{\circ}$-174$^{\circ}$
                         &   \cite{Spiekermann2007a} \\  
\noalign{\smallskip}\hline
\end{tabular}
\end{center}
\end{table}


\subsection{$\mathrm{Au_nGe_{12}}$ Clusters}
\label{sec:geosweeps}

Now we examine linked $\mathrm{Ge_6Au_nGe_{6}}$ configurations 
to compare with $\mathrm{Au_3Ge_{18}}$.  
There are innumerable combinations of 
two octahedral $\mathrm{Ge_6}$ cages 
connected to one or more Au atoms, 
so we design input geometries to 
test the linearity of Ge-Au-Ge links and 
accommodate the symmetries of the $\mathrm{Ge_6}$ cages.
A basic question is when will there be sufficient 
electrons such that the cages maintain the octahedral symmetry 
of the Wade-like Ge$_6^{2-}$ clusters and also bond together
through the gold interconnections.
We ran CG relaxations of many such input geometries, 
but illustrate just a few of the most interesting cases 
which have similar initial $\mathrm{Ge_6}$ orientations 
as those described in Section~\ref{sec:Ge-clusters}, 
now with intercluster Au atoms. 

With just one Au atom, we examined orientations 
very similar to the CG structures I$_0$ and II$_0$ 
with an Au atom added. 
For both of these orientations, 
the axes of the octahedra are aligned 
with an Au atom at their midpoint.  
The input structure for I$_0$ produced 
I$_1$, where a total of four Au-Ge links could form 
in the equatorial planes of the two octahedra.  
Likewise II$_0$ and II$_1$ shared input structures where 
the added Au atom in 
in II$_1$ created two Au-Ge links 
in a line connecting the apexes of the two octahedra.  
With two Au atoms, 
many possible interlinked combinations exist, 
but stable isomers resulted from 
placing both Au atoms
in the common equatorial plane of the two octahedra,
creating four Au-Ge links.  
With three Au atoms,  
we focused on canted octahedra, 
similar to structure III$_0$ [Fig.~\ref{fig:cgmdge12}(c)], 
to create a nine-atom prism much like that 
in $\mathrm{Au_3Ge_{18}}$.  
 
Such initialization procedures requires an exploration of geometry phase space 
to ensure our CG relaxations do not find highly metastable states.  
In order to sample many possible configurations we use 
a ``geometry sweep'' much like the approach used to find 
the groundstate of crystal structures in DFT methods.  
In general, a sweep was generated by using as a variable the 
\textit{initial} separation $d\mathrm{_{init}}$ of Au from Ge atoms 
to which it was interconnected.  
For a given initial cluster orientation, 
independent CG relaxations of the cluster were run 
for a series of Au-Ge separations, 
yielding many isomers to compare, each with
a relaxed geometry and binding energy.  
A single geometry sweep explored one input orientation 
with 15-20 different Au-Ge separations 
while setting initial Ge-Ge distances at 3.0~\AA.
The CG relaxations returned dense clusters with high binding energies for short Au-Ge distances,
dissociated clusters with low binding energies at long Au-Ge distances, and
intermediate clusters which retained two $\mathrm{Ge_6}$ cages connected by Au atoms.  
Generally within a broad range of $d\mathrm{_{init}}$, these intermediate clusters are 
nearly identical after CG relaxation.  
From these intermediate isomers we present those with relatively high binding energies. 

Figure~\ref{fig:sweep} shows two geometry sweeps for $\mathrm{AuGe_{12}}$, 
where each point represents a separate isomer. 
Each sweep produces compact clusters at $d\mathrm{_{init}} < 2.5$~\AA, 
intermediate clusters at $2.5 < d\mathrm{_{init}} < 5.0$~\AA, 
and dissociated clusters at higher separations.
Sweep 1, with four Au-Ge interconnects,   
produces I$_1$ in the range $2.5 < d\mathrm{_{init}} < 5.0$~\AA, 
and yields low E$_B$ structure V$_1$ at $d\mathrm{_{init}} = 2.5$~\AA. 
Sweep 2, with two Au-Ge interconnects, 
produces many isomers: 
II$_1$-III$_1$ observed where 
$3.0 < d\mathrm{_{init}} < 3.6$~\AA, 
IV$_1$ in the range 
$3.8 < d\mathrm{_{init}} < 4.8$~\AA, 
and VI$_1$ and VII$_1$ 
at $2.2 <d\mathrm{_{init}} < 3.0$~\AA.  
We show the geometries of 
I$_1$ and II$_1$ in Fig.~\ref{fig:auge12_cg_md} and 
III$_1$-VII$_1$ in Fig.~\ref{fig:supplemental_structures}.  
The geometry differences within a sweep are small 
and the overall range of $E_B$/atom 
between isomers is less than 0.1 eV, 
indicating near degeneracy of the clusters.  
Interesting differences include those of II$_1$ and III$_1$, 
where III$_1$ is twisted about its long axis compared with II$_1$. 
$V_1$ is considerably more octahedral that $I_1$. 

\begin{figure}
\resizebox{0.5\textwidth}{!}{%
  \includegraphics{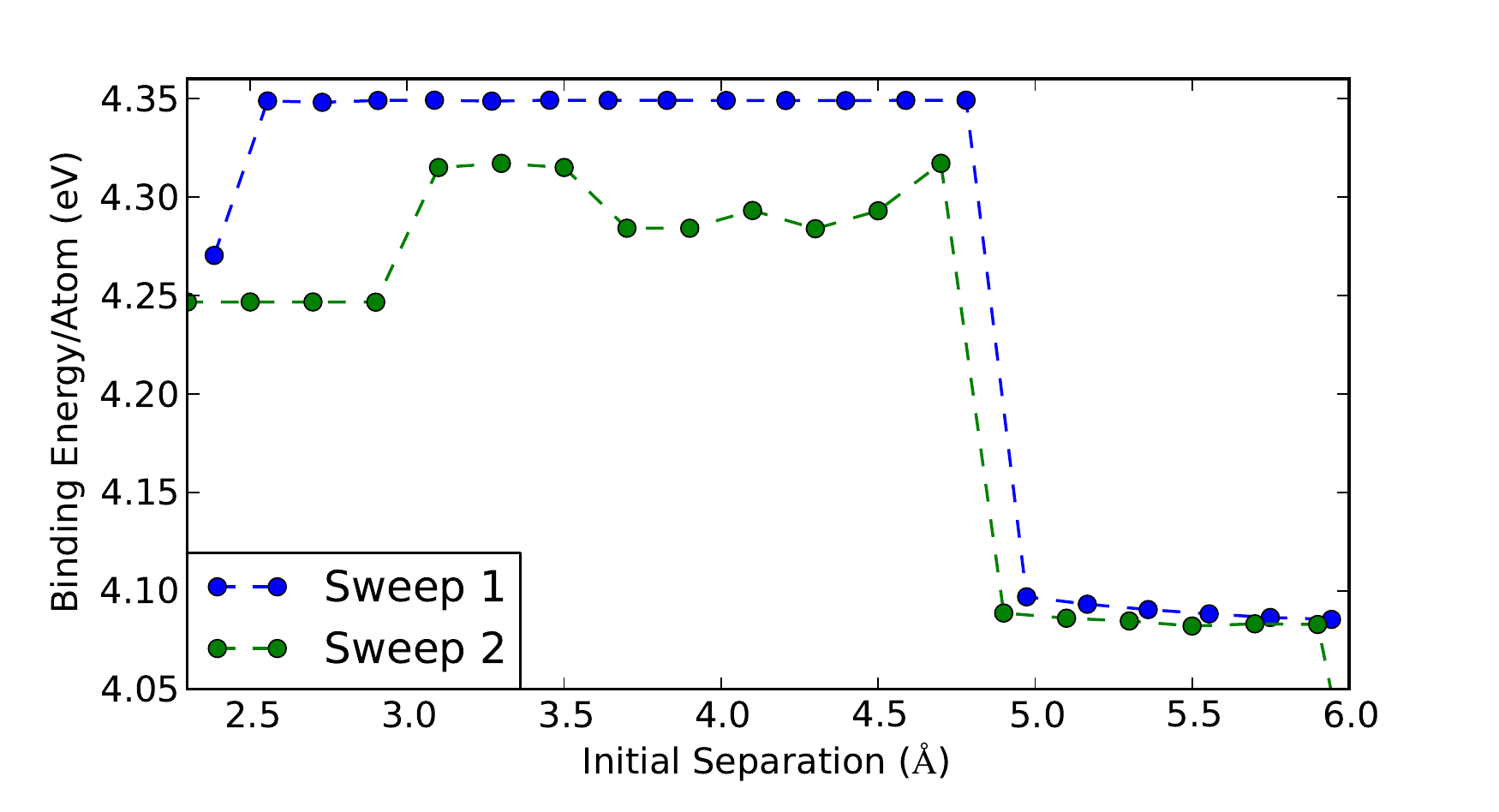}
}
\caption{Geometry sweep for clusters $\mathrm{AuGe_{12}}$: 
where the CG-relaxed geometry and corresponding $E_B$ (eV)
are shown as a function of initial Au-Ge separation.  
The input structures of Sweep 1 and 2 are 
described in the text.  
}
\label{fig:sweep}   
\end{figure}  

\begin{table*}
  \begin{center}
\caption{Calculated average skeletal distances (\AA) 
  and deviations from octahedrality (\AA)
  of the CG-relaxed clusters 
  $\mathrm{Au_nGe_{12}}$ shown in Figs.~\ref{fig:auge12_cg_md} and 
  \ref{fig:supplemental_structures}.
  Also shown, 
  ranges for bond lengths $d_{G}$, $d_{AG}$, $d_{A}$, and  
  values for the angles $\theta$ 
  connecting Au to the two cages and energy $E_B$ (eV).  }
\label{tab:cgge12aun}      
\begin{tabular}{llllllll}
\hline\noalign{\smallskip}
$\mathrm{AuGe_{12}}$ &$\bar{d}_G$ & $\sigma_{d_G}$ & $d_{AG}$ & & $d_G$ & $\theta$ & $E_B$ \\
\noalign{\smallskip}\hline\noalign{\smallskip}
 I$_1$    & 2.731 & 0.242 & 2.66  & & 2.58-3.29 & $104^{\circ}$& 4.349 \\ 
 II$_1$  &  2.663 & 0.127 & 2.46  & & 2.53-2.86 & $180^{\circ}$ & 4.317 \\ 
       III$_1$ & 2.664 & 0.127 & 2.465 & & 2.51-2.88 & $180^{\circ}$ & 4.315 \\ 
       IV$_1$  & 2.689 & 0.189 & 2.49  & & 2.55-2.94 & $180^{\circ}$ & 4.284 \\ 
       V$_1$    & 2.703 & 0.086 & 2.53  & & 2.54-2.77 & $114^{\circ}$ & 4.270 \\ 
       VI$_1$   & 2.673 & 0.124 & 2.49  & & 2.57-2.83 & $180^{\circ}$ & 4.266 \\ 
       VII$_1$  & 2.652 & 0.028 & 2.46  & & 2.63-2.69 & $180^{\circ}$ & 4.247 \\  
\hline\noalign{\smallskip}
$\mathrm{Au_2Ge_{12}}$ &$\bar{d}_G$ & $\sigma_{d_G}$ & $d_{AG}$ &  $d_A$ & d$_{G}$ & $\theta$ & $E_B$ \\
\noalign{\smallskip}\hline\noalign{\smallskip}
 I$_2$   & 2.735 & 0.343 & 2.46-2.51 & 2.89 & 2.50-3.76  
        & 171.4$^{\circ}$-176$^{\circ}$ & 4.292 \\
        II$_2$   & 2.837 & 0.443 & 2.42-2.52 & 2.90 & 2.48-3.83 
        & 117-176$^{\circ}$ & 4.291 \\
        III$_2$ & 2.709 & 0.232 & 2.52-2.64 & 2.81 & 2.52-3.14 
        & 172.5$^{\circ}$ & 4.256 \\ 
        IV$_2$ & 2.654 & 0.063  & 2.46 & 4.09 & 2.57-2.72 
        & 148$^{\circ}$ & 4.241 \\
\hline\noalign{\smallskip}
$\mathrm{Au_3Ge_{12}}$ & $\bar{d}_G$ & $\sigma_{d_G}$ & $d_{AG}$ & $d_A$ & $d_G$ & $\theta$ & $E_B$ \\
\noalign{\smallskip}\hline\noalign{\smallskip}
           I$_3$  & 2.653 & 0.063 & 2.46-2.50 & 4.15 & 2.56-2.71 & $130-138^{\circ}$ & 4.191 \\ 
           II$_3$ & 2.661 & 0.048 & 2.49-2.81 & 3.45-3.59 & 2.58-2.73 & $109-175^{\circ}$ & 4.187\\ 
           III$_3$ & 2.685 & 0.056 & 2.48-2.52 & 3.06 & 2.62-2.76 & $136-151^{\circ}$ & 4.181\\ 
           IV$_3$ & 2.683 & 0.142 & 2.50-2.54 & 2.97 & 2.51-2.93 & $112-139^{\circ}$ & 4.145 \\ 
           V$_3$  & 2.760 & 0.356 & 2.51-2.69 & 2.78-2.94 & 2.52-3.74 & $174^{\circ}$ & 4.135 \\
\noalign{\smallskip}\hline
\end{tabular}
\end{center}
\end{table*}

\begin{figure}
\resizebox{0.5\textwidth}{!}{%
  \includegraphics{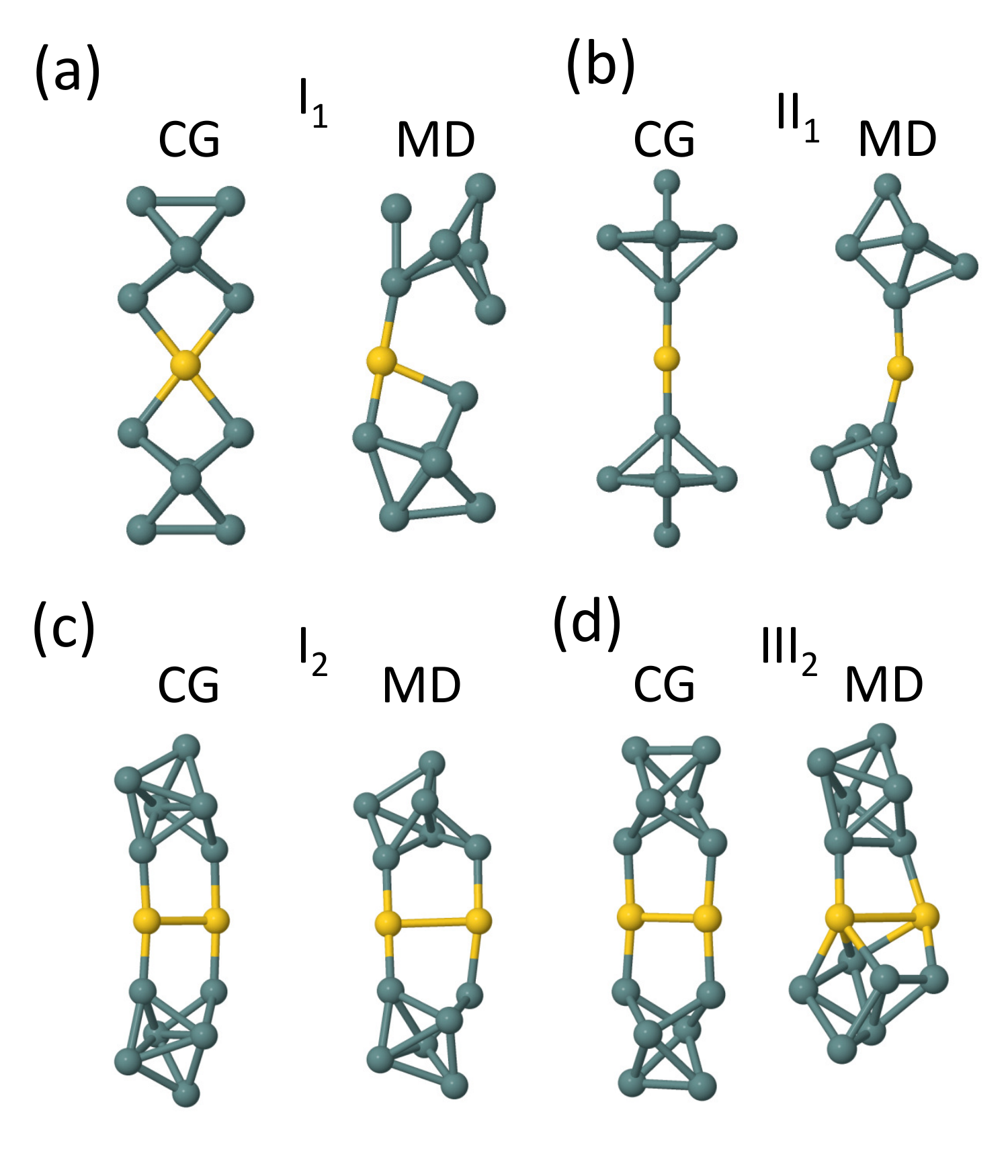}
}

\caption{Examples of $\mathrm{Au_nGe_{12}}$ clusters 
  after CG relaxation (left) and post-MD (right).   
  (a) $n=1$, structure I$_1$; 
  (b) $n=1$, structure II$_1$; 
  (c) $n=2$, structure I$_2$; 
  (d) $n=2$, structure III$_2$. 
  Au atoms are gold and Ge are gray.  
  }
\label{fig:auge12_cg_md}      
\end{figure}

In each group $\mathrm{Au_nGe_{12}}$, $n=1-3$, 
we identified a few 
highest binding energy structures for careful study 
and present geometry and energy data  
of their CG relaxations 
in Table~\ref{tab:cgge12aun} and those for post-MD in Table~\ref{tab:mdaunge12}.   
We examine clusters which maintain a 
recognizable pair of $\mathrm{Ge_6}$ cages 
throughout a CG relaxation. 
Differences in energies 
between the highest and lowest 
in each group was small, 
typically less that 0.1 eV. 
There are no significant trends in the deviation from octahedrality 
${\sigma_{d_G}}$ with binding energy.  
Ge-Au-Ge angles $\theta$ vary from straight to bent 
(e.g., there is one structure, II$_3$, 
in the $\mathrm{Au_3Ge_{12}}$ series that has a near tetrahedral angle).  
Due to the large number of CG-relaxed $\mathrm{Au_nGe_{12}}$ 
structures summarized in this table,
we show only four representative $n=1$ and $n=2$ structures in Fig.~\ref{fig:auge12_cg_md}.
Since we observe no stable 
neutral \AuGe{3}{12} clusters, for $n=3$ 
we show the CG-relaxed and 
post-MD geometries of $\mathrm{Au_2Ge_{12}^{4-}}$ 
in Fig.~\ref{fig:au2ge12_-4_cg_md} 
and that of $\mathrm{Au_3Ge_{12}^{2-}}$ 
in Fig.~\ref{fig:au3ge12_-2_cg_md}, which will be discussed in further detail below.
The remainder of the
structures listed in this table can be found in 
Fig.~\ref{fig:supplemental_structures}. 

\begin{figure}
\resizebox{0.5\textwidth}{!}{%
  \includegraphics{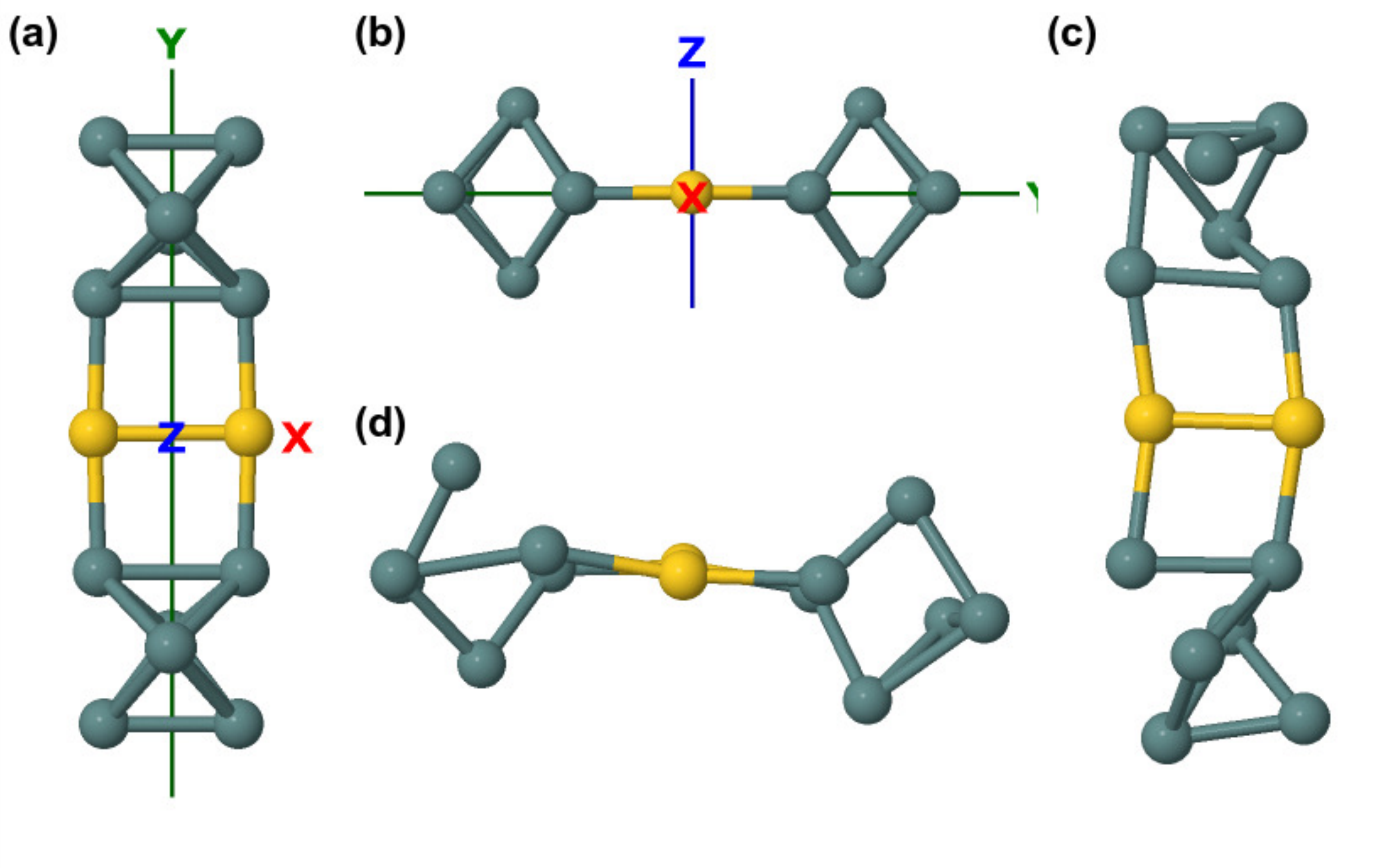}
}

\caption{The $\mathrm{Au_2Ge_{12}^{4-}}$ cluster 
  with 
  (a-b) CG relaxation and (c-d) post-MD.   
  }
\label{fig:au2ge12_-4_cg_md}      
\end{figure}
\begin{figure}
\resizebox{0.5\textwidth}{!}{%
  \includegraphics{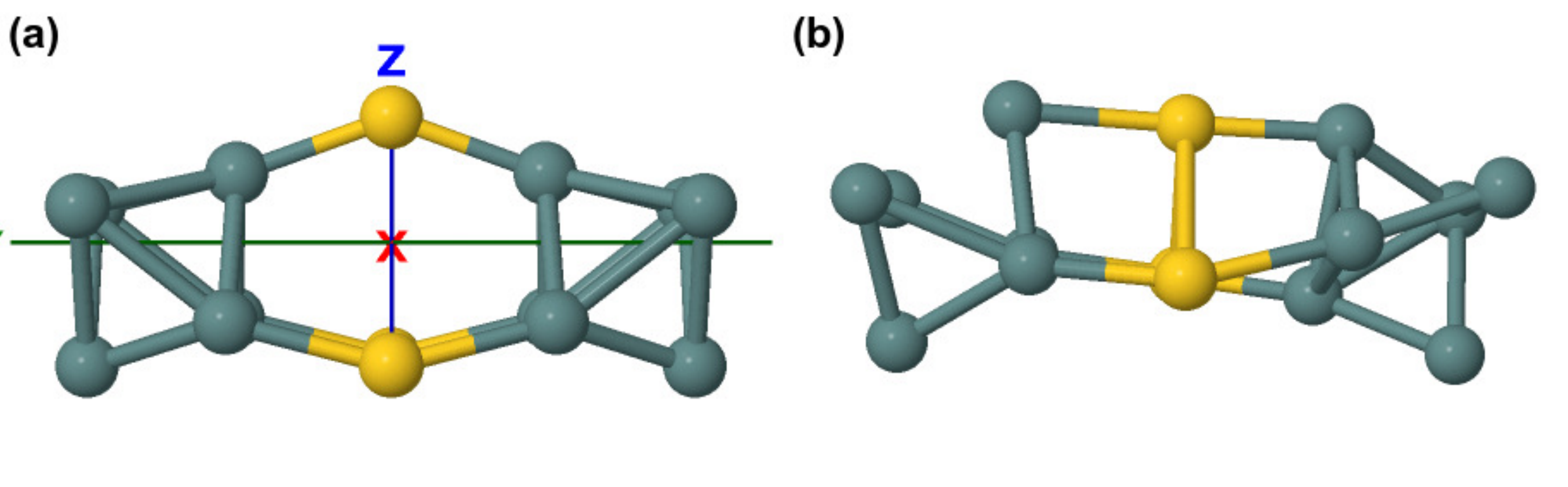}
}

\caption{The $\mathrm{Au_3Ge_{12}^{2-}}$ cluster 
  with 
  (a) CG relaxation and (b) post-MD. 
  The three Au atoms form a triangle 
  in the x-z plane, 
  and the lower, back Au atom is hidden from view.  
  }
\label{fig:au3ge12_-2_cg_md}      
\end{figure}

Figure \ref{fig:auge12_cg_md} compares $n=1$
isomers (a) I$_1$ and (b) II$_1$ 
that have respectively four and two Ge-Au-Ge intercluster links
after CG relaxation.  
The octahedra of I$_1$ are more deformed than II$_1$, 
with $\sigma_{d_G}$ larger by 0.12~\AA, 
the Au-Ge bonds are longer in  I$_1$ by 0.2~\AA, 
and $E_B$ is greater by approximately 0.03 eV.  
Post-MD, we see I$_1$ (a) breaks an Au-Ge intercluster bond, 
while II$_1$ (b) maintains both: merely twisting about its Ge-Au-Ge 
intercluster link so $\theta_{\mathrm{Ge-Au-Ge}}$ remains close to linear. 
When comparing the two predominant isomers 
of $\mathrm{AuGe_{12}}$, 
the four Au-Ge intercluster bonds of I$_1$ 
are less stable than the two of II$_1$.

For $n=2$, Fig.~\ref{fig:auge12_cg_md} compares the
isomers (c) I$_2$ and (d) III$_2$.  
After CG relaxation, I$_2$ is less symmetric than III$_2$ ($C_{2v}$ vs. $D_{2h}$) 
which coincides with reduced octahedrality in 
I$_2$ ($\sigma_{d_G}$ = 0.343~\AA).  
II$_2$ (Fig.~\ref{fig:supplemental_structures}) is similar to I$_2$, 
with more exaggerated asymmetry ($\sigma_{d_G}$ = 0.443~\AA), 
while IV$_2$ (Fig.~\ref{fig:supplemental_structures}) 
is similar to III$_2$ with a strongly bent 
Ge-Au-Ge link ($172.5^{\circ}$ vs $148^{\circ}$)
and a widely separated Au-Au distance (4.09~\AA).
The overall symmetry change between the isomers of $\mathrm{Au_2Ge_{12}}$ 
appears to have direct consequences for the overall cluster stability.
The stability for $n=2$ under MD simulation appears 
to rely more on symmetry than number of intercluster links: 
the asymmetric I$_{2}$ and II$_{2}$ structures 
are stable under MD while the 
relatively symmetric  III$_{2}$ and IV$_{2}$ structures are not 
although each have four linear Au-Ge intercluster links.  
This is consistent with the asymmetric 
coupling of the $\mathrm{Ge_9}$ clusters 
in $\mathrm{[Au_3Ge_{18}]^{5-}}$.  
We see that $\mathrm{Au_3Ge_{12}}$ does not 
readily form linear stable Ge-Au-Ge links, 
whether the relative \GeM{6} orientations are 
symmetric or asymmetric.

\begin{table*}
\caption{Results for the 
  calculated average skeletal distances 
  and deviations from octahedrality (\AA) 
  of the $\mathrm{Au_nGe_{12}}$ clusters after MD runs, 
  where a blank in a $\sigma_{d_G}$ column indicates a complete disordering 
  of the octahedra.  
  Also shown, values for the angles $\theta$ 
  connecting Au to the two cages and energy, $E_B$ (eV).  
  Structures are shown in Figs.~\ref{fig:auge12_cg_md} and 
  \ref{fig:supplemental_structures}.  }
\label{tab:mdaunge12}     
\begin{center}
\begin{tabular}{l l l l l l l l l}
\hline\noalign{\smallskip}{\smallskip}
 $\mathrm{AuGe_{12}}$ &  & $d_{AG}$ & $\theta$  &  
 $({\bar{d}}_G)_1$ & ${(\sigma_{d_G})}_1$ & $(\bar{d}_G)_2$ &
 ${(\sigma_{d_G})}_2$ & $d_G$ \\
\noalign{\smallskip}\hline\noalign{\smallskip}
    I$_1$  & & 2.46, 2.76 &  $166^{\circ}$ & 
    2.74 & 0.32 & 2.79 & 0.40 & 2.36-3.51 \\    
    II$_1$ & & 2.45,2.60 & $163^{\circ}$ & 
    2.74 & 0.40 & 2.73 & 0.23 & 2.42-3.95 \\
\hline\noalign{\smallskip}
  $\mathrm{Au_2Ge_{12}}$ & $d_A$ & $d_{AG}$ & $\theta$  &  
  $({\bar{d}}_G)_1$ & ${(\sigma_{d_G})}_1$ & $(\bar{d}_G)_2$ &
  ${(\sigma_{d_G})}_2$ & $d_G$ \\
\noalign{\smallskip}
\hline\noalign{\smallskip}{\smallskip}
        I$_2$   & 3.20 & 2.49-2.66 & 163$^{\circ}$,177$^{\circ}$ & 
        2.779 & 0.379 & 2.743 & 0.309 & 2.53-3.73 \\
        II$_2$  & 3.03 & 2.47-2.73 & 173$^{\circ}$ & 
        2.774  & 0.318 & 2.812 & 0.409 & 2.43-3.87\\
        III$_2$ & 3.42 & 2.41-2.84 & 102$^{\circ}$-157$^{\circ}$ & 
        2.696 & 0.298 & --- & --- & 2.39-3.56\\
        IV$_2$  & 3.27 & 2.38-2.73 & 125$^{\circ}$-163$^{\circ}$ & 
        2.726 & 0.251 & --- & --- & 2.49-3.41\\
\noalign{\smallskip}
\hline\noalign{\smallskip}{\smallskip}
 $\mathrm{Au_3Ge_{12}}$ & $d_A$ & $d_{AG}$ & $\theta$  &  
$({\bar{d}}_G)_1$ & ${(\sigma_{d_G})}_1$ & $(\bar{d}_G)_2$ &
${(\sigma_{d_G})}_2$ & $d_G$ \\
\noalign{\smallskip}
\hline\noalign{\smallskip}{\smallskip}
 I$_3$  & 3.04 & 2.46-3.04 &  $75^{\circ}-169^{\circ}$ & 
 ---  & ---   & --- & --- & 2.52-3.21  \\
 II$_3$ & 3.60-3.71 & 2.48-3.01 & $100^{\circ}-175^{\circ}$  & 
 2.68 & 0.11 & --- & --- & 2.54-2.99  \\
 III$_3$ & 3.25-3.68 &  2.43-2.91 & $96^{\circ}-165^{\circ}$ & 
 2.81 & 0.28 & --- & --- & 2.54-3.53  \\
 IV$_3$ & 2.75  & 2.45-2.89 &  $127^{\circ}-143^{\circ}$ & 
 2.73 & 0.19 &  --- & --- & 2.37-3.29 \\
 V$_3$ & 2.77 &  2.48-2.84 &  $150^{\circ}$ & 
 2.81 & 0.37 &  --- & --- & 2.51-3.83 \\
\noalign{\smallskip}\hline
\end{tabular}
\end{center}
\end{table*}

Charge influences cluster shape, as dictated by Wade's rules.
We consider here the effect of a charge $q=2$ and $q=4$  
on some isomers of $\mathrm{Au_nGe_{12}}^{q-}$.
To obtain these results, we ran identical geometry sweeps 
to our neutral charge calculations, with charge $q=2$ and $q=4$ added to the cluster.  

Table~\ref{tab:charge_table} lists 
two stable isomers that were found for 
$\mathrm{Au_2Ge_{12}}^{2-}$.  
The primary difference is in the linearity of the 
Ge-Au-Ge links and the degree of octahedrality, 
with lower energy isomer having 
$\theta = 148.5^{\circ}$, like the neutral IV$_2$, 
and $\sigma_{d_G} = 0.05$~\AA.  
Nearly degenerate is the isomer with 
$\theta = 178.5^{\circ}$, like the neutral III$_2$.  
When the charge is increased to 
$q=4$, only the linear-link isomer is observed, 
which is also stable under MD simulation.  
Likewise we observed a 
stable isomer of $\mathrm{Au_3Ge_{12}}^{q-}$. 
with both $q=2$ and $q=4$.  
While the CG-relaxed $\mathrm{Au_3Ge_{12}}^{2-}$ 
still has bent Ge-Au-Ge links ($\theta = 140^{\circ}-149^{\circ}$), 
the post-MD structure has one nearly linear bond ($\theta = 178^{\circ}$), 
and the bridging structure resembles a prism.  
This structure type was 
simply not observed in the neutral cluster.

In contrast to the neutrally charged isomers, 
we observe linear links in both $\mathrm{Au_2Ge_{12}^{q-}}$, 
$q=2$ and $q=4$ with $D_{2h}$ symmetry (as III$_2$) 
that are stable under MD simulation. 
Moreover we observe stable $\mathrm{Au_3Ge_{12}^{q-}}$ 
clusters.  
While  a more complete exploration of 
both value of $q$ and possible linking structures could be performed,  
this work confirms that additional charge 
stabilizes linear Ge-Au-Ge cluster interconnections.

\begin{table*}
\begin{center}
\caption{Calculated properties of the CG-relaxed clusters 
  found in geometry sweeps 
  for charged clusters $\mathrm{Au_nGe_{12}}^{q-}$.    
  Also shown, the properties for these
  clusters post-MD.}
\label{tab:charge_table}      
\begin{tabular}{l l l l l l l}
\noalign{\smallskip}\hline\noalign{\smallskip}
        \multicolumn{7}{c}{CG Properties}\\
        \hline\noalign{\smallskip}
           & $\bar{d}_G$ & $\sigma_{d_G}$ & $d_{AG}$ 
           & $d_A$ & $d_{G}$ & $\theta$ \\
         Au$_2$Ge$_{12}^{2-}$ (1) & 2.66 & 0.05 & 2.49 & 3.92 & 2.56-2.73 & $148.5^{\circ}$\\
        Au$_2$Ge$_{12}^{2-}$ (2) & 2.70 & 0.15 & 2.53 & 2.84 & 2.53-2.95 & $178.5^{\circ}$\\
          Au$_2$Ge$_{12}^{4-}$ & 2.72 & 0.09 & 2.59 & 2.89 & 2.51-2.84 & $176^{\circ}$\\
          Au$_3$Ge$_{12}^{2-}$ & 2.67 & 0.06 & 2.58-2.75 & 3.14-4.0 & 2.51-2.84 & $140^{\circ}-149^{\circ}$\\
          Au$_3$Ge$_{12}^{4-}$ & 2.73 & 0.13 & 2.76 & 3.23-3.72 & 2.59-2.93 & $155^{\circ}$\\
\noalign{\smallskip}\hline\noalign{\smallskip}
    \multicolumn{7}{c}{MD Properties}\\
\hline\noalign{\smallskip}    
       & $\bar{d}_G$ & $\sigma_{d_G}$ & $d_{AG}$ 
       & $d_A$ & $d_{G}$ & $\theta$ \\
     Au$_2$Ge$_{12}^{2-}$ (1) & 2.67 & 0.11 & 2.46-2.48 & 3.80 & 2.35-4.05 & $140^{\circ}$,$157^{\circ}$\\
     Au$_2$Ge$_{12}^{2-}$ (2) & 2.72 & 0.23 & 2.38-2.60 & 3.82 & 2.40-2.88 & $149^{\circ}$,$157^{\circ}$\\        
     Au$_2$Ge$_{12}^{4-}$ & 2.79 & 0.17 & 2.62-2.70 & 2.76 & 2.51-3.26 
            & $161^{\circ}-164^{\circ}$\\
       Au$_3$Ge$_{12}^{2-}$ & 2.75 & 0.15 & 2.37-2.73 & 3.22-3.96 & 2.43-2.99 & $146^{\circ}-178^{\circ}$\\
       Au$_3$Ge$_{12}^{4-}$ & 2.82 & 0.24 & 2.58 & 2.93-4.16 & 2.58-3.39 & $135^{\circ}-171^{\circ}$\\
\noalign{\smallskip}\hline\noalign{\smallskip} 
\end{tabular}
\end{center}
\end{table*}

The DOS of the \AuGe{n}{12} clusters generally resembles 
that of \AuGe{3}{18} where the Ge $4p$ orbitals dominate 
just below $E_F$ with some overlap with the Au $5d$ states.  
The Ge $4p$ and Au $5d$ states 
of I$_1$ have less overlap than II$_1$.  
The DOS of I$_2$ and III$_2$ are qualitatively similar, 
but the additional symmetry of III$_2$ promotes $E_F$ 
above the energy gap of I$_2$, indicating an electronic instability.   
We see the most overlap between 
Ge and Au orbitals in \AuGe{3}{12} 
due to the greater magnitude of the Au states, 
but it is clear from MD that this does not contribute to a stability of the 
$\mathrm{Ge_6Au_nGe_6}$ motif when $n=3$.  
We present the HOMO and LUMO of selected \AuGe{n}{12} clusters 
in Figs.~\ref{fig:au1ge12_I_HL}-\ref{fig:au3ge12_I_HL}
where delocalized bonding 
is apparent in the LUMO states of $n=1,3$ 
and the HOMO states in $n=2$ 
\AuGe{n}{12} clusters.  

In COHP analysis, we use the structures 
shown in Fig.~\ref{fig:auge12_cg_md} 
to compare the properties of  
stable CG-relaxed 
$\mathrm{Au_nGe_{12}}$ clusters: 
I$_1$ and II$_1$ in $\mathrm{AuGe_{12}}$ and 
I$_2$ and III$_2$ in $\mathrm{Au_2Ge_{12}}$.
We show their COHP interactions in three figures:
Ge-Ge in Fig.~\ref{fig:aunge12_ge-ge_COHP}; 
Au-Ge in Fig.~\ref{fig:aunge12_au-ge_COHP}; and
Au-Au in Fig.~\ref{fig:aunge12_au-au_COHP}.
Generally we see that the primary indicator 
of stability is that of the Au-Ge interaction.  

In $\mathrm{AuGe_{12}}$ 
the overall similarities between I$_1$ and II$_1$ 
in the the Ge-Ge interaction 
of are apparent in 
Fig.~\ref{fig:aunge12_ge-ge_COHP} (a-b) respectively 
where a single, large bonding peak appears below $E_F$  
at $E-E_F \approx -2.0$~eV.
However I$_1$ shifts 
from Ge-Ge bonding to antibonding states 
at $E_F$ while II$_1$ has a small antibonding state in Ge-Ge at $E_F$.  
In the Au-Ge interaction 
[Fig.~\ref{fig:aunge12_au-ge_COHP} (a-b)], 
I$_1$ has antibonding states at $E_F$, 
while II$_1$ 
has no Au-Ge state at $E_F$.  
Instead $E_F$ is centered 
between occupied bonding and empty antibonding states.  
Given our observations during MD simulation 
that I$_1$ is less stable, 
this indicates the Ge-Ge antibonding state 
arises from intercluster interaction, 
like that of I$_0$, 
but does not break the structure.  
The antibonding Au-Ge state at $E_F$ of I$_1$
is likely to cause 
the Au-Ge bond breaking observed in the post-MD cluster, 
whereas those in II$_1$, with no such antibonding interaction, 
merely twist, but remain bonded.  

The Ge-Ge COHP curves of 
$\mathrm{Au_2Ge_{12}}$ I$_2$ and III$_2$ 
are qualitatively similar to one another
[Figs.~\ref{fig:aunge12_ge-ge_COHP}(c,d)].  
Neither cluster has antibonding states below $E_F$, 
and the primary difference is the  
shift in the location of $E_F$ from I$_2$ to III$_2$.  
Significantly, while the Au-Ge interactions of I$_2$ 
have no antibonding character below $E_F$ 
[Fig.~\ref{fig:aunge12_au-ge_COHP}(c)], 
there is an antibonding interaction 
in III$_2$ [Fig.~\ref{fig:aunge12_au-ge_COHP}(d)], 
again confirming that antibonding states in Au-Ge below $E_F$ 
are correlated with cluster instability.  
In examining COHP curves of $\mathrm{Au_3Ge_{12}}$, 
there are as many as three antibonding states below $E_F$ 
in Au-Ge interactions, 
thus it is unsurprising 
that no neutral stable clusters were observed.  

We see no overall stability trends in 
the Au-Au COHP results. 
In Fig.~\ref{fig:aunge12_au-au_COHP} 
we show the Au-Au COHP results of 
(a) I$_2$ and (b) III$_2$.  
In both cases, 
there are significant anti-bonding MO's below $E_F$.  
In terms of antibonding states, 
the Au-Au COHP for $\mathrm{Au_2Ge_{12}}$.   
is consistent with that observed for the charged $\mathrm{Au_3Ge_{18}^{5-}}$.
To further examine this antibonding state 
the electronic structure analysis 
should account for 
aurophilic Au-Au interactions as discussed in 
Refs.~\cite{Spiekermann2007,Schmidbaur2008}.  

\begin{figure}
\resizebox{0.5\textwidth}{!}{%
  \includegraphics{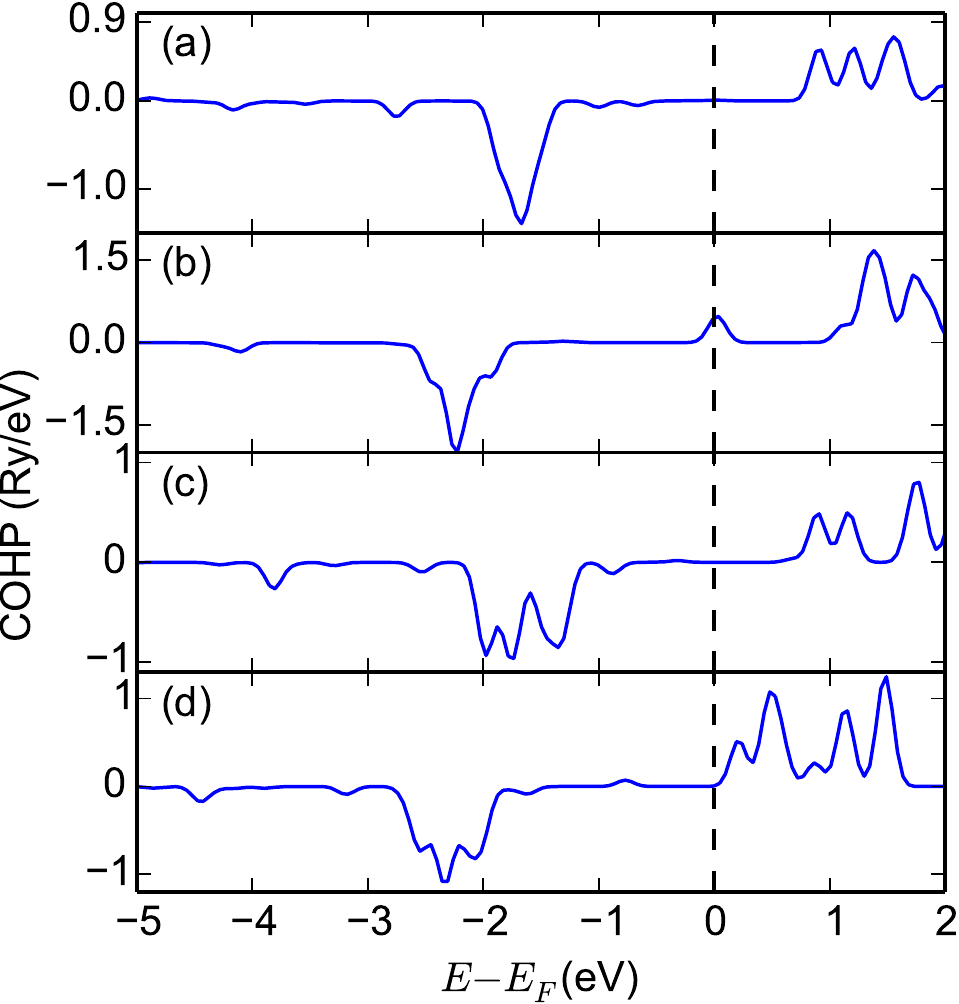}
}
\caption{COHP for Ge-Ge bonds for $\mathrm{Au_nGe_{12}}$
where (a) I$_1$, (b) II$_1$, (c) I$_2$, (d) III$_2$}
\label{fig:aunge12_ge-ge_COHP}      
\end{figure}
%
\begin{figure}
\resizebox{0.5\textwidth}{!}{%
  \includegraphics{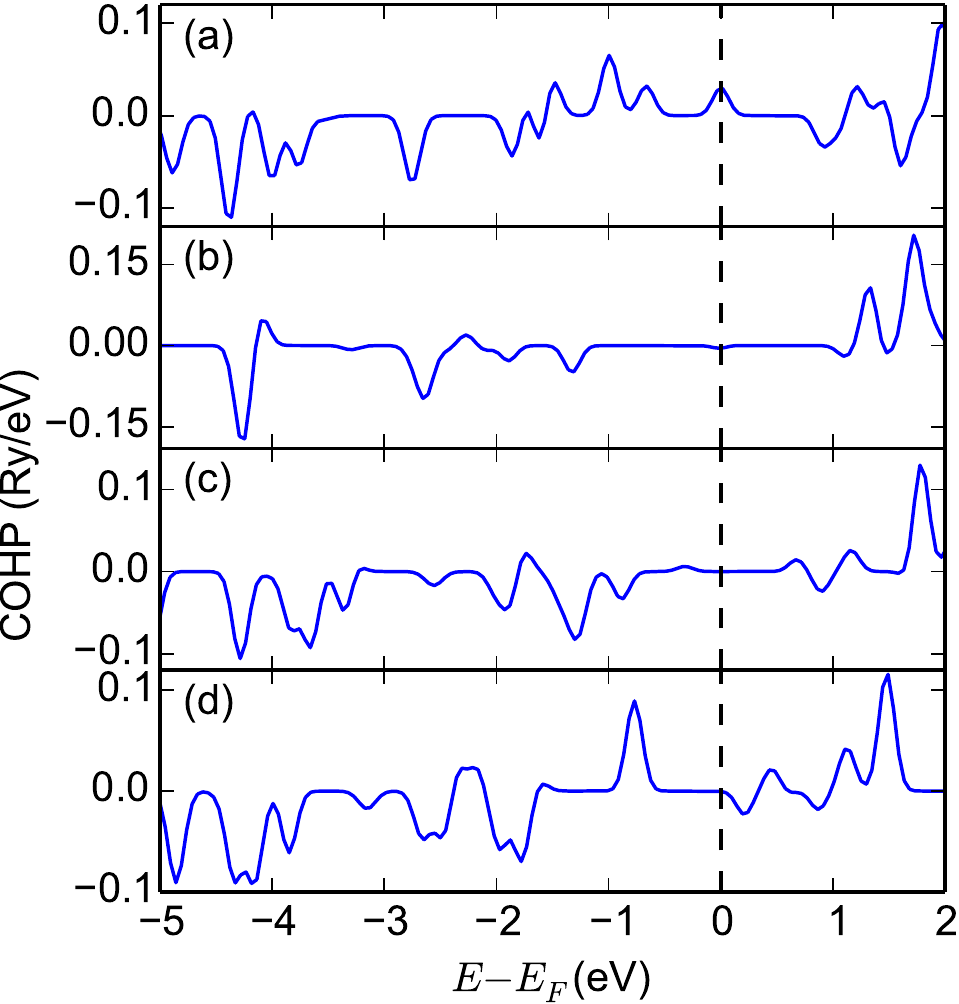}
}
\caption{COHP for Au-Ge bonds for $\mathrm{Au_nGe_{12}}$ 
where 
(a) I$_1$, (b) II$_1$, (c) I$_2$, (d) III$_2$.}
\label{fig:aunge12_au-ge_COHP}       
\end{figure}
%
\begin{figure}
\resizebox{0.5\textwidth}{!}{%
  \includegraphics{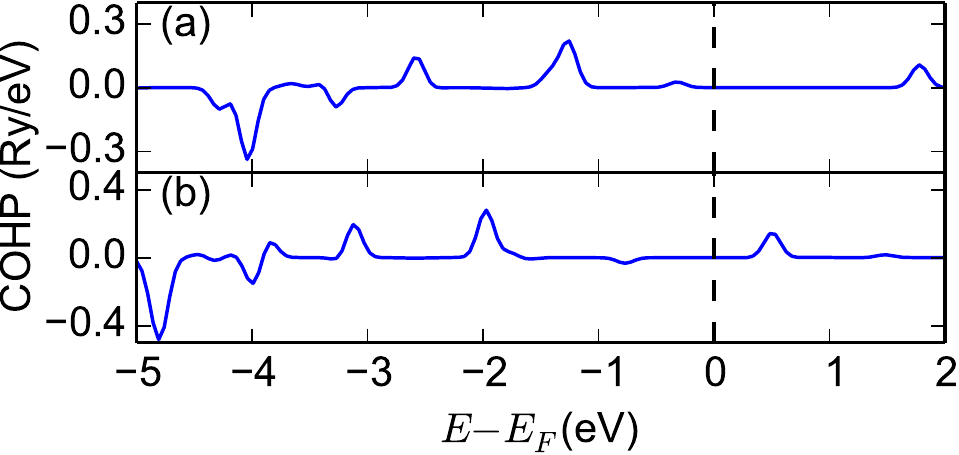}
}
\caption{COHP for Au-Au bonds for $\mathrm{Au_2Ge_{12}}$.
where (a) I$_2$, (b) III$_2$.}
\label{fig:aunge12_au-au_COHP}      
\end{figure}
%

\subsection{Extended $_2^\infty[\mathrm{Au_2Ge_{6}}]$ Structures}
\label{sec:compounds}

To extend our exploration of the ability of  $\mathrm{Ge_6}$ cages 
to bond with Au atoms, 
we consider crystalline Au-Ge structures built with this geometry.  
The precedent studies include Ref.~\cite{Karttunen2010}, 
where 1D, 2D, and 3D structures formed of $\mathrm{Ge_9}$ cages 
were examined.  
We show the hypothetical structure $_2^\infty[\mathrm{Au_2Ge_{6}}]$ 
in the inset of Fig.~\ref{fig:extended} 
which represents 
the isomer observed with binding energy maximum 
at $d\mathrm{_{init}} = 2.6$~\AA.  
Here, Au atoms are placed in the positions 
of the radial non-bonding orbitals of 
isolated  $\mathrm{[Ge_6]^{2-}}$ clusters in the $xy$ plane.   
Following the techniques of Sec.~\ref{sec:geosweeps}, 
we vary the initial Au-Ge distance, perform CG minimization, 
and find the binding energy and relaxed geometry 
(Fig.~\ref{fig:extended}).  
The relaxed structure features a subunit of 
a $\mathrm{Ge_6}$ octahedron with four radial Au-Ge bonds. 
Each subunit is 
spaced closely to form a 2D crystal structure including 
Au-Au bonds. 
The binding energy has a peak at $d_{\mathrm{init}}=2.6$~\AA, 
where $d_{\mathrm{init}}$ is the initial Au-Ge bondlength.  
The average bond found in the Ge cages 
is $\bar{d}_G=2.72 \pm 0.06$~\AA, 
showing that the bonds are elongated compared to the isolated clusters, 
but the cages are close to octahedral.  
Ge-Ge bonds range from $d_G= 2.62-2.77$~\AA, 
which includes those 
in the two-dimensional plane forming 
a square with lengths of approximately 2.65~\AA~and 
the out-of-plane bonds are 2.77~\AA.  
The Au-Ge bond lengths $d_A=2.42$~\AA~are short 
compared with the clusters studied in Sec.~\ref{sec:Ge-clusters} 
and 
the Au-Au bond is $d_A=2.85$~\AA.  
The Ge-Au-Ge linking angle is 117$^\circ$.  
We explored other stoichiometries, 
such as $_1^\infty[\mathrm{Au_1Ge_{6}}]$, 
$_1^\infty[\mathrm{Au_2Ge_{6}}]$, and 
$_2^\infty[\mathrm{Au_4Ge_{6}}]$, 
and observed considerably less smooth 
E$_B$ vs. $d_{\mathrm{init}}$ curves, 
indicating these stoichiometries prefer dissociation 
or non-octahedral geometries 
in the input structures we examined.  
Again we see that 
a Au-Ge ratio of 1:6 seems to support greater 
overall stability in Au-Ge systems.
%
\begin{figure}
\resizebox{0.5\textwidth}{!}{%
  \includegraphics{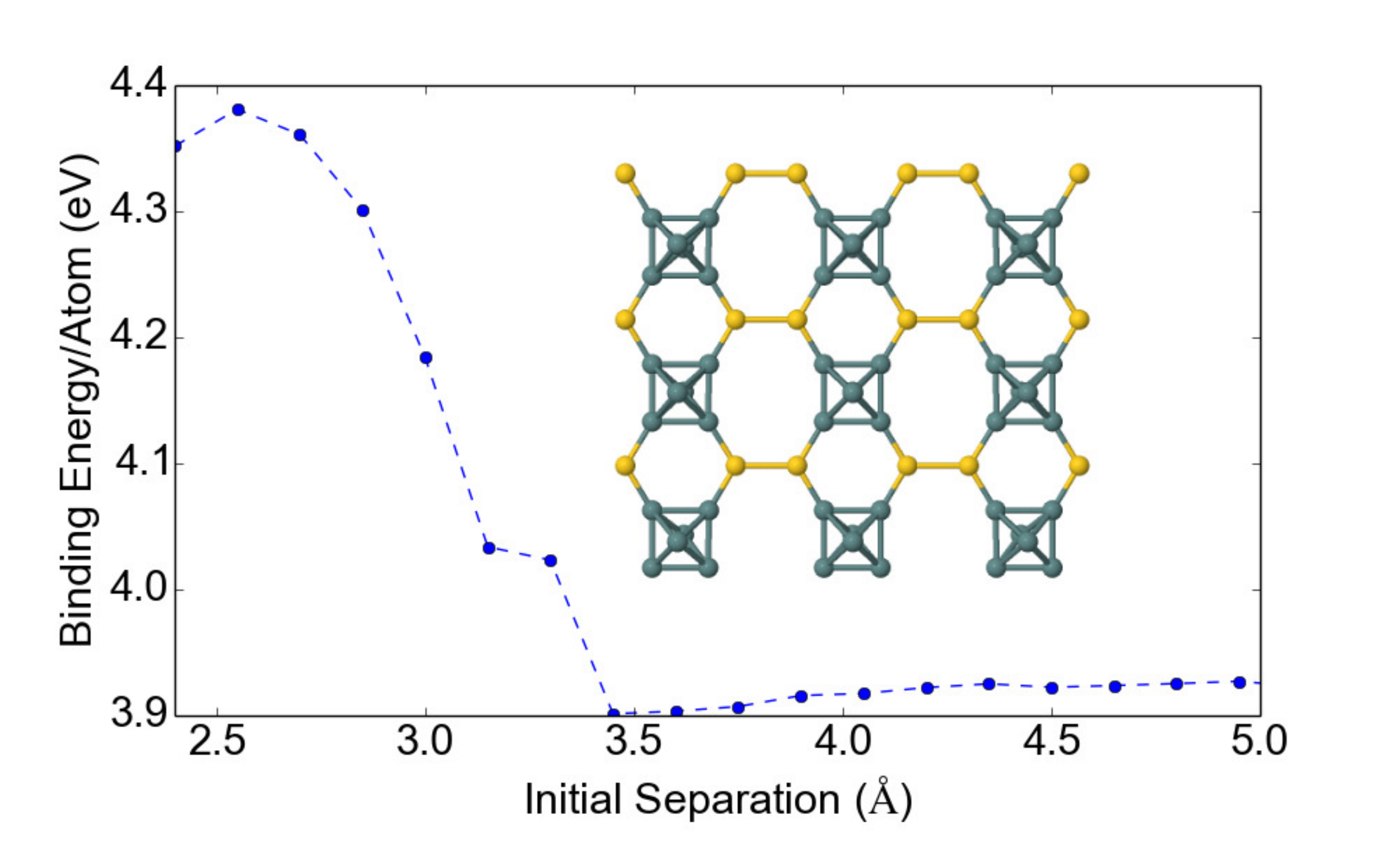}
}
\caption{Binding energy $E_B$ (eV) 
  versus $d_{\mathrm{init}}$ (\AA) of extended structure 
  $_2^\infty[\mathrm{Au_2Ge_{6}}]$.  
  The inset shows 
  the geometry of this local $E_B$ maxima
  found with CG minimization for $d_{\mathrm{init}}=2.6$~\AA.   }
\label{fig:extended}       
\end{figure}

\section{Discussion and Conclusions}
\label{sec:discussion}

Beyond its well known properties as 
a covalently bonded $sp^3$ semiconductor, 
the versatility in the 
Ge electronic structure allows it to 
form cage-like structures with extended bond lengths.  
Taken in isolation, cages $\mathrm{Ge_6}$ and $\mathrm{Ge_9}$ 
need additional electrons to stabilize their deltahedral shapes; 
two electrons being sufficient for the former, 
and four for the latter, consistent with Wade's Rules.  
By using DFT methods, 
we have examined the stability of 
combinations of 
octahedral $\mathrm{Ge_6}$
cages in various forms.  
In examining linking structures between two such cages, 
we found that two bonds form and stabilize $\mathrm{Ge_{12}}$,
especially when  
the two clusters are aligned symmetrically 
with Ge-Ge bonds connecting the two equatorial planes of the octahedra.  
Other orientations are less stable, 
as can be seen using COHP analysis or 
by running brief MD simulations to examine how the cluster evolves.  

By combining Au with Ge, new possibilities emerge.  
It is tantalizing that through linking to 
a triangle of Au atoms, 
deltahedral $\mathrm{Ge_9}$ cages stabilize 
in the charged \\ $\mathrm{[Au_3Ge_{18}]^{5-}}$.  
In analogy, 
this paper addressed whether $\mathrm{Ge_6}$ cages 
could be linked into $\mathrm{Au_nGe_{12}}$ clusters 
for $n=1-3$.  
We observe stable isomers of readily form in 
$\mathrm{AuGe_{12}}$ and $\mathrm{Au_2Ge_{12}}$, 
but not $\mathrm{Au_3Ge_{12}}$.  
This appears largely due to the nature of the 
overlap in molecular orbitals at the highest occupied level: 
it is relatively simple to favor the radially directed 
Au-Ge bonds in $n=1-2$ systems, 
and difficult in $n=3$.  
We also observe additional stability provided 
by asymmetric combinations of $n=2$ systems 
and charge in both $n=2-3$ systems.  
Our conclusion is that stable Au-Ge clusters 
or perhaps even extended structures 
are most likely to be found if 1-2 Au atoms are used to 
interconnect the equatorial planes of the octahedra, 
assuming external ligands or large spacing ions were employed 
to counterbalance ionicity.  
This suggests that geometry and charge state play 
a larger role in stable intercluster links than 
chemical species 
consistent with Wade's rules.  

Remaining still as a theoretical challenge 
are such problems as 
is the kinetic nature of Au's role in 
crystallizing Ge \cite{Tan1992} 
or forming diamond-structure nanowires \cite{Wang2002}.  
Recent work \cite{Gamalski2012} in nanowire growth 
observes metastable crystalline AuGe catalysts 
which may further inform which geometries are favored in 
these nano-sized clusters.  
Our brief MD studies were not sufficient to make progress in this challenge.  
But perhaps understanding better Au-Ge bonds in existing compounds 
should be the center of future studies 
of this fascinating pair of elements.


\section*{Acknowledgements}
This research was supported in part by the assistance and resources of
the Notre Dame Center for Research Computing, 
particularly Paul Brenner and Timothy Stitt. 

\bibliography{au_ge_ref}

\beginsupplement
\newpage
\section{Supplementary Material}
In Fig.~\ref{fig:au3ge18_HL} 
we show the Highest Occupied Molecular Orbital 
(HOMO) and Lowest Unoccupied Molecular Orbital (LUMO) 
for the CG-relaxed cluster  $\mathrm{Au_3Ge_{18}}$.  
These figures clearly show the asymmetry of the two $\mathrm{Ge_9}$ cages. 
In the LUMO pictures, the right cage
shows extended $\pi$-like hybridization 
connecting into triangle of Au atoms.  
The nodal structure of the orbitals 
for both cages changes from HOMO to LUMO.
Note, COHP for 
the Au-Au bonds of this cluster 
is shown in Fig.~\ref{fig:au3ge18_COHP}.

In Tables 4 and 6, a number of clusters  
$\mathrm{Au_nGe_{12}}$, $n=1,2,3$,  are listed and described.  
Four figures were picked as being representative examples for the main paper 
and are shown in Fig. 8 (specifically, I$_1$, II$_1$, I$_2$, and III$_2$).  
The remainder of those clusters are shown in Fig. S2.
Clusters  III$_1$ through VII$_1$ are shown only after CG minimization
since the isomers have the same overall symmetry and number of bonds 
as I$_1$ and II$_1$, 
and thus exhibit 
similar behavior under MD simulation. 
The rest are shown after both CG minimization and MD simulation.

We show HOMO and LUMO states for five selected 
CG-relaxed clusters of  $\mathrm{Au_nGe_{12}}$.
COHP analysis is also shown for four of these five clusters in Sec. 3.7.
In Fig. S3, we show two orientations each for  
$\mathrm{AuGe_{12}}$, cluster I$_1$.
Referring back to Fig. 8(a), we see that it is highly symmetric.  
An interesting feature 
is the spherical orbital 
centered on the Au atom, seen in (a) and (c),
connecting only loosely to the extended $\pi$-like orbitals of the two cages.  
In the LUMO, (b) and (d), 
the center Au atom is not directly connected to the cages.  
As discussed in Sec. 3.7, this cluster is less stable than cluster II$_1$, 
we show the post-MD 
rearrangement and symmetry breaking of I$_1$ in Fig. 8(a).  

In Fig. S4, 
we show the orbitals after CG minimization 
for the more stable $n=1$ cluster II$_1$.  
Cluster II$_1$ is again highly symmetric [Fig. 8(b)] 
but interestingly, the HOMO state is degenerate and shows
very localized electronic density on the cages, 
with very minimal interaction with 
the central Au atom in this orbital.
The LUMO state, Fig. S4(c), 
has electronic structure for the Au bridge site 
that is similar to that shown in Fig. S3(b),
that is, an extended $\pi$-like orbital 
that does not include the central Au atom.  
The orientations of the cages 
relative to the bridging Au are different for I$_1$ and II$_1$, 
as seen when comparing 
Figs. S3 to S4 or 
comparing the CG picture for I$_1$ of 
Fig. 8(a) to the one for II$_1$ in Fig. 8(b).

As Figs. S3 and S4 corresponded to Figs. 8(a) and (b), 
in Figs. S5 and S6, 
we show the HOMO and LUMO states for $n=2$, 
which correspond to
the images of CG-relaxed structures in 
Figs. 8(c) and (d).  
As discussed in Sec. 3.7, 
cluster I$_2$, Figs. S5 and 8(c),
is more stable than cluster III$_2$, 
Figs. S6 and 8(d).  
The cage orbitals for the HOMO state for
$n=2$ are less extended than for $n=1$.  
Because of the relative tilting of the two cages for I$_2$, 
the two Au atoms participate differently in the bonding in the HOMO,
Fig. S5.  
In contrast, 
for III$_2$ shown in Fig. S6, 
both Au atoms are equally involved in bonding 
to each other in the HOMO, and not at all in the LUMO.

As seen in Fig. S2, 
all $n=3$ clusters change significantly after a MD run.  
In Fig. S7, 
we show HOMO and LUMO for cluster I$_3$ to 
compare these with its CG-relaxed geometry [Fig. S2(h)].
The tilting of the two cages causes 
two of the three Au atoms to participate differently 
in the bonding in the HOMO and in the LUMO states.  
It is most interesting to compare this with 
the HOMO and LUMO states for $\mathrm{Au_3Ge_{18}}$, Fig. S1.  
Here we see that the
cage electrons in 
$\mathrm{Au_3Ge_{12}}$ are much more localized than in $\mathrm{Au_3Ge_{18}}$.

\begin{figure*}
\resizebox{0.95\textwidth}{!}{%
  \includegraphics{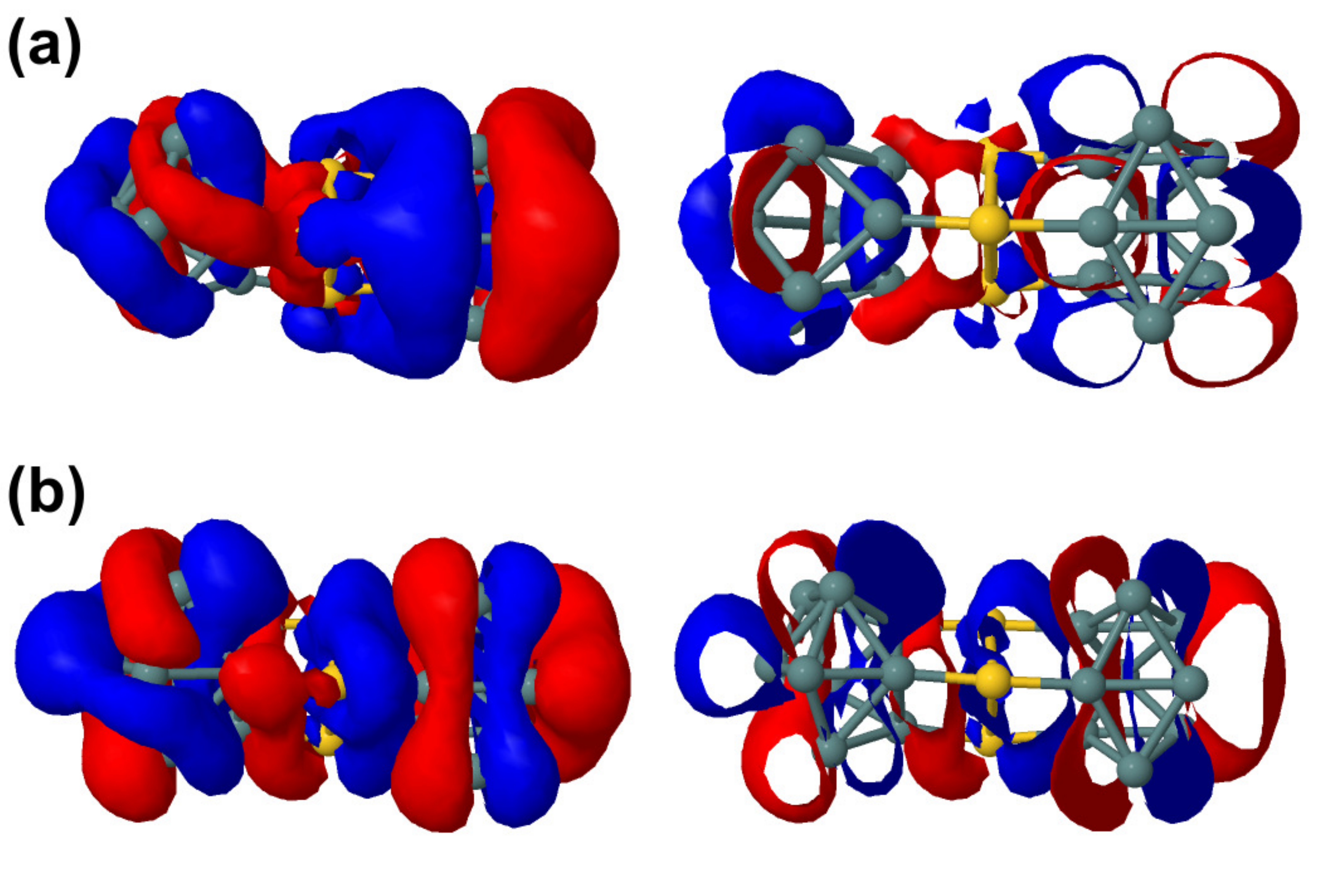}
}
\caption{
(a) HOMO and (b) LUMO of neutral $\mathrm{[Au_3Ge_{18}]}$ 
where the right column is a slice of the 3D orbitals shown on the left.  
} 
\label{fig:au3ge18_HL}       
\end{figure*}

\begin{figure*}
\resizebox{7.0in}{!}{%
  \includegraphics{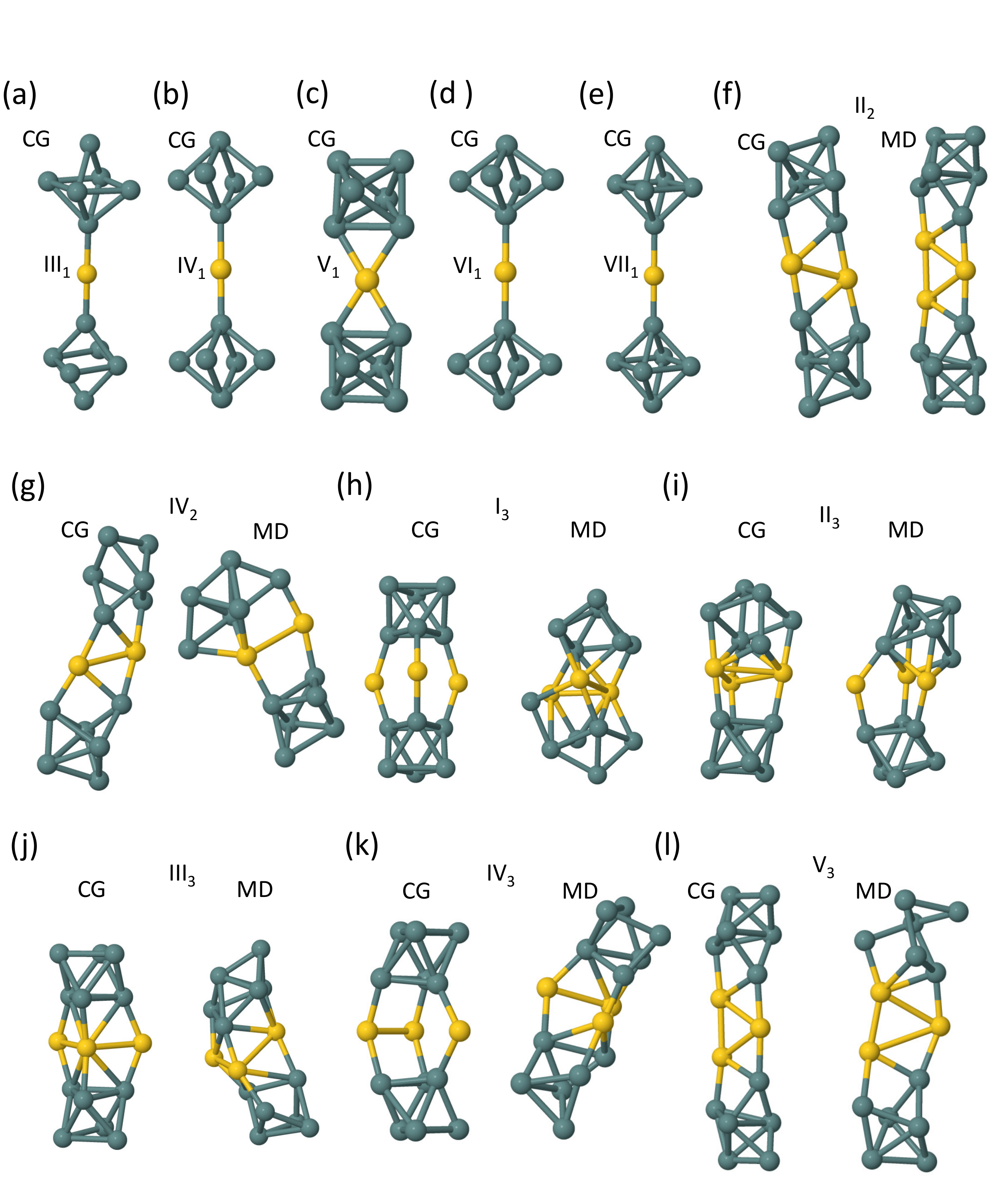}
}
\caption{%
Clusters of $\mathrm{Au_nGe_{12}}$: 
(a)-(e), $n=1$, III$_1$, IV$_1$, V$_1$, VI$_1$, VII$_1$;
(f) and (g), $n=2$,  II$_2$ and (b) IV$_2$; 
(h) - (l):, $n=3$,  I$_3$, II$_3$, III$_3$, IV$_3$, V$_3$.
For $n=1$, 
clusters are shown only for after CG minimization.     
Cluster IV$_1$ only differs from VI$_1$ due to 
a smaller horizontal cross section.  
For $n=2,3$, clusters are shown after 
CG minimization and also after a MD run.
} 
\label{fig:supplemental_structures}.
\end{figure*}

\begin{figure*}
\resizebox{7.0in}{!}{%
  \includegraphics{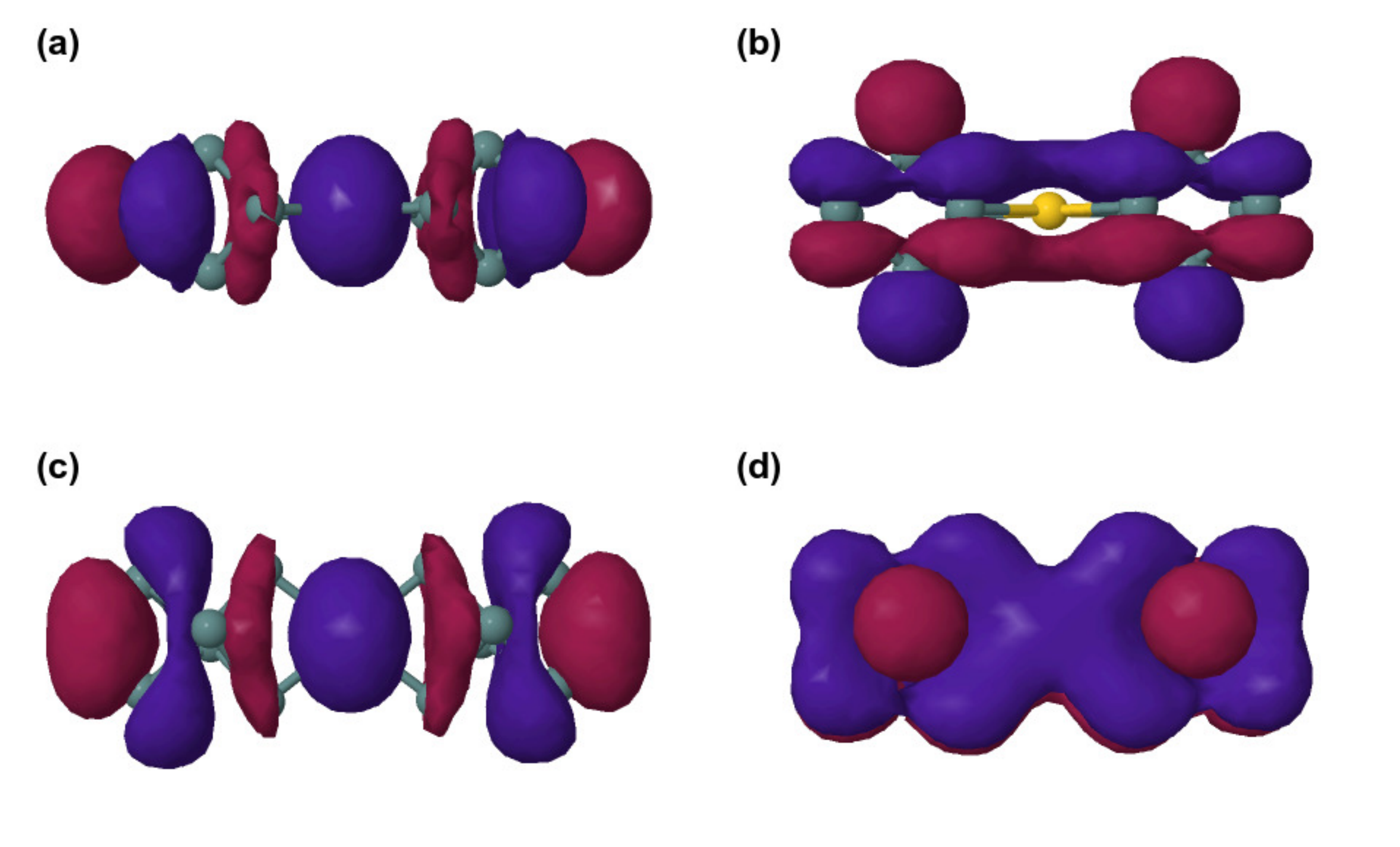}
}
\caption{
(a,c) HOMO and (b,d) LUMO of neutral \AuGe{}{12} I$_1$.
} 
\label{fig:au1ge12_I_HL}       
\end{figure*}

\begin{figure*}
\resizebox{7.0in}{!}{%
  \includegraphics{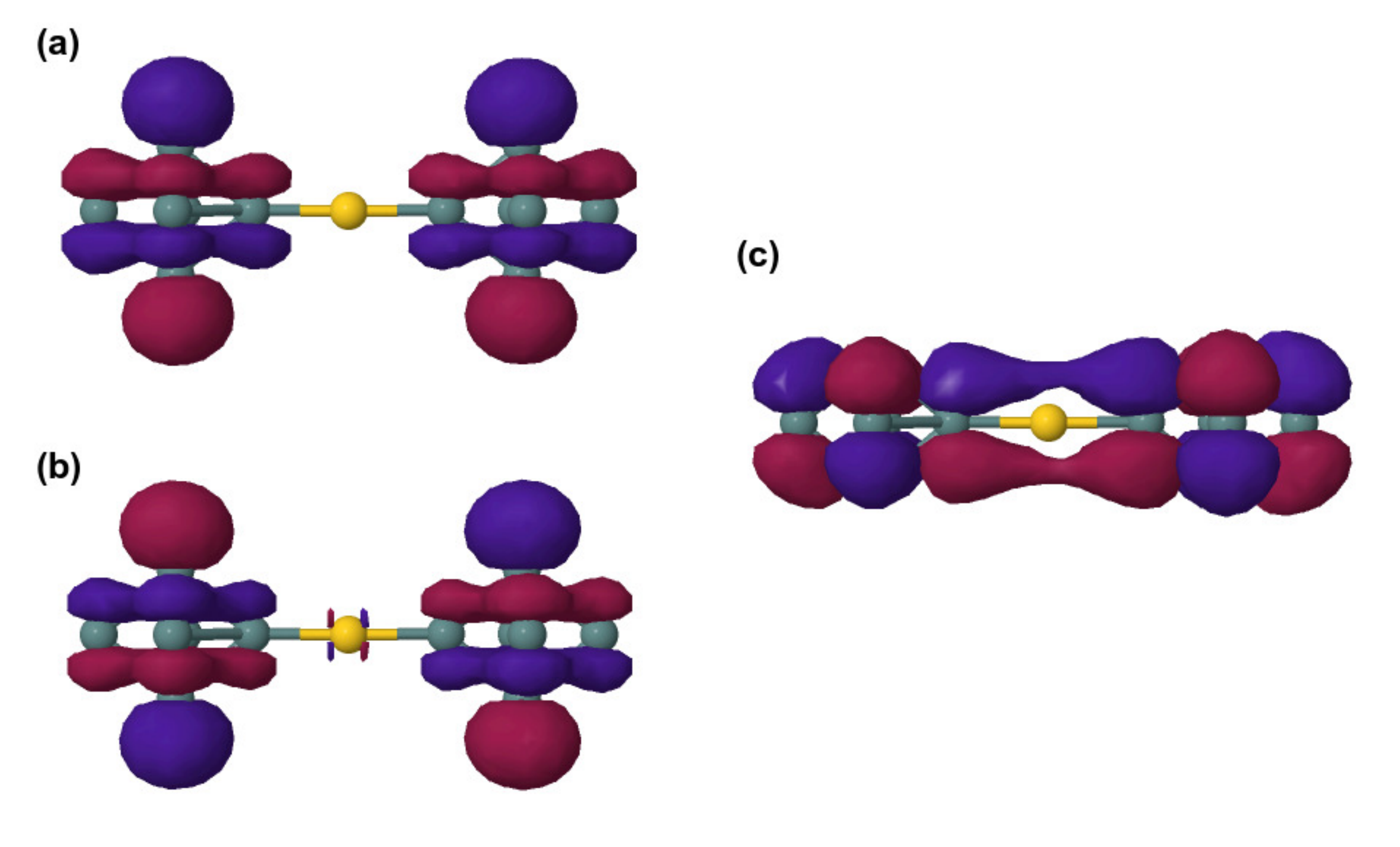}
}
\caption{
(a,b) Degenerate HOMO and (c) LUMO of neutral \AuGe{}{12} II$_1$.
} 
\label{fig:au1ge12_II_HL}       
\end{figure*}

\begin{figure*}
\resizebox{7.0in}{!}{%
  \includegraphics{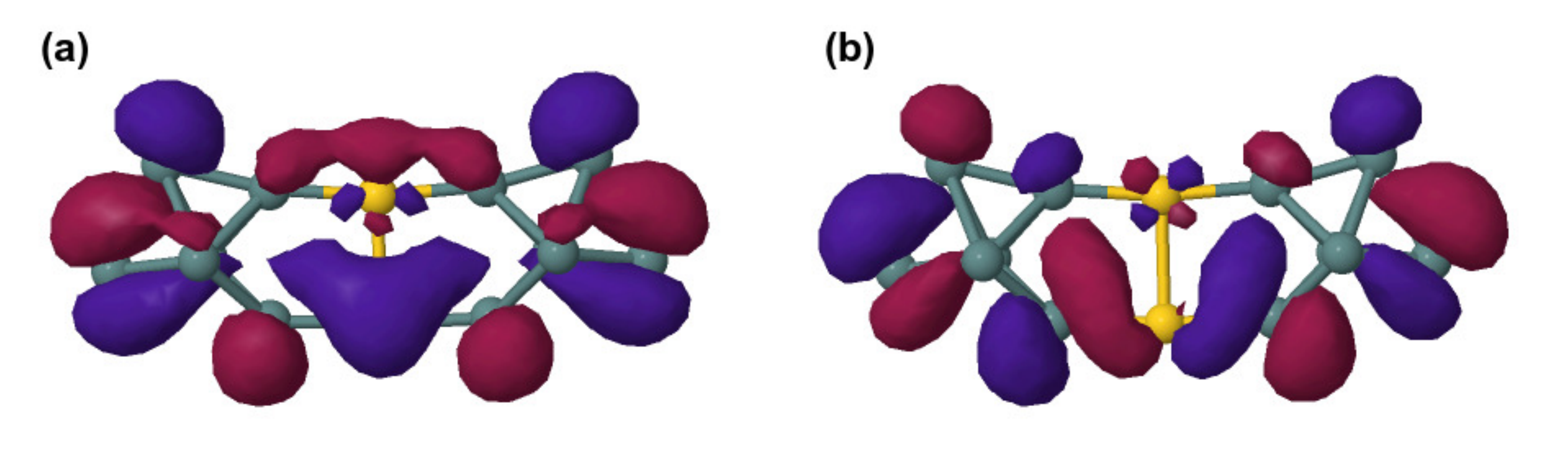}
}
\caption{
(a) HOMO states and (b) LUMO of neutral \AuGe{2}{12} I$_2$.
} 
\label{fig:au2ge12_I_HL}       
\end{figure*}

\begin{figure*}
\resizebox{7.0in}{!}{%
  \includegraphics{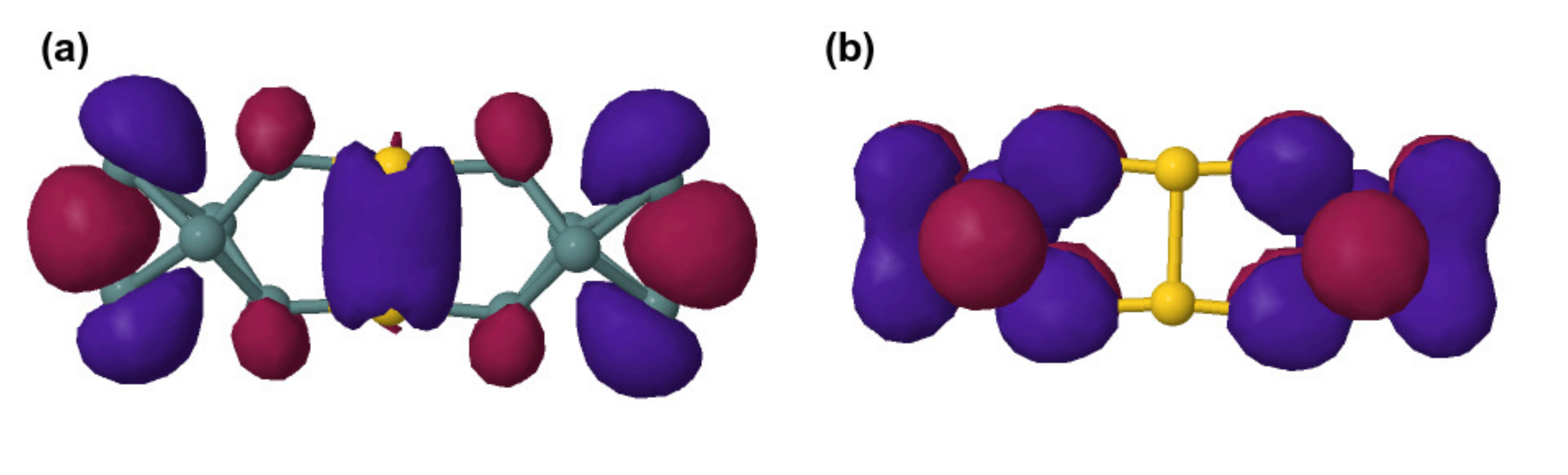}
}
\caption{
(a) HOMO and (b) LUMO of neutral \AuGe{2}{12} III$_2$.
} 
\label{fig:au2ge12_III_HL}       
\end{figure*}

\begin{figure*}
\resizebox{7.0in}{!}{%
  \includegraphics{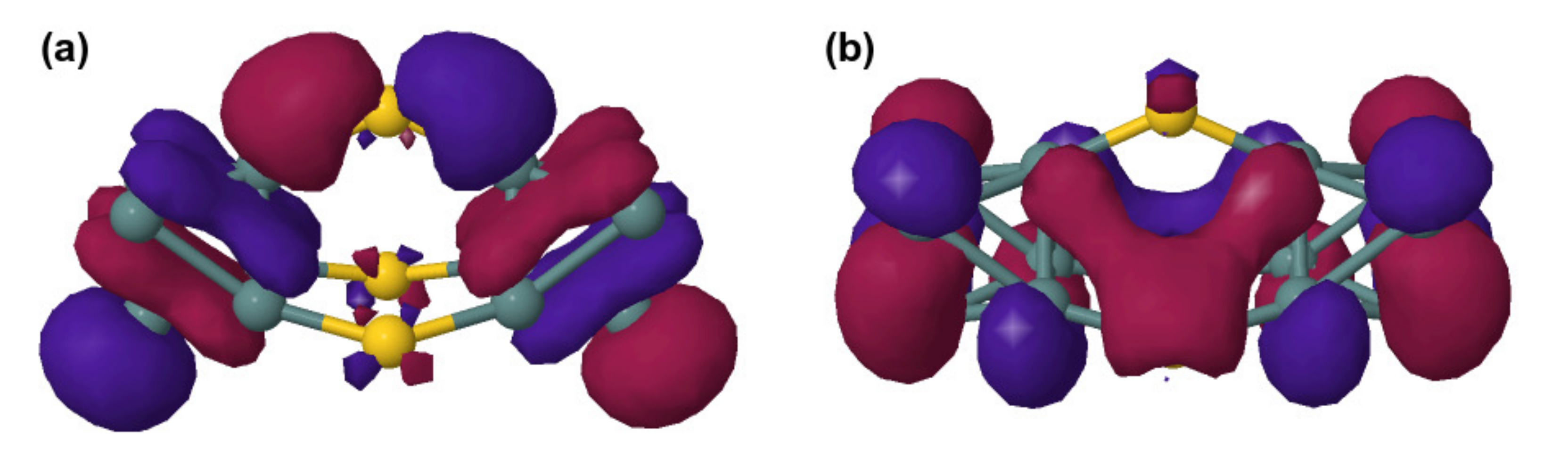}
}
\caption{
(a) HOMO and (b) LUMO of neutral \AuGe{3}{12} I$_3$.
} 
\label{fig:au3ge12_I_HL}       
\end{figure*}

\end{document}